\RequirePackage{lineno}
\documentclass[prb,twocolumn,floatfix,showpacs,superscriptaddress]{revtex4-1}
\usepackage[pdftex,plainpages=false,colorlinks=true,linkcolor=blue, citecolor=blue, urlcolor=blue]{hyperref}
\usepackage{amsfonts}
\usepackage{amsmath}
\usepackage{amssymb}
\usepackage{graphicx}
\usepackage{natbib}
\DeclareMathAlphabet\mathbfcal{OMS}{cmsy}{b}{n} 

\begin{document}

\title{Pseudogap and superconductivity in two-dimensional doped charge-transfer insulators}
\author{L. Fratino}
\affiliation{Department of Physics, Royal Holloway, University of London, Egham, Surrey, UK, TW20 0EX}
\author{P. S\'emon}
\affiliation{D\'epartement de physique and Regroupement qu\'eb\'equois sur les mat\'eriaux de pointe, Universit\'e de Sherbrooke, Sherbrooke, Qu\'ebec, Canada J1K 2R1}
\author{G. Sordi}
\affiliation{Department of Physics, Royal Holloway, University of London, Egham, Surrey, UK, TW20 0EX}
\author{A.-M. S. Tremblay}
\affiliation{D\'epartement de physique and Regroupement qu\'eb\'equois sur les mat\'eriaux de pointe, Universit\'e de Sherbrooke, Sherbrooke, Qu\'ebec, Canada J1K 2R1}
\affiliation{Canadian Institute for Advanced Research, Toronto, Ontario, Canada, M5G 1Z8}

\date{\today}

\begin{abstract}
High-temperature superconductivity emerges upon doping a state of matter that is insulating because of interactions. A widely studied model considers one orbital per CuO$_2$ unit cell on a square lattice with a strong intra-orbital repulsion that leads to a so-called Mott-Hubbard insulator. Here we solve a model that takes into account, within each unit cell, two oxygen orbitals where there is no electron-electron repulsion and a copper orbital with strong electron-electron repulsion. The insulating phase is a so-called  charge-transfer insulator, not a Mott-Hubbard insulator. Using cluster dynamical mean-field theory with continuous-time quantum Monte Carlo as an impurity solver and 12 atoms per cluster, we report the normal and superconducting phase diagram of this model as a function of doping, interaction strength and temperature. As expected, the three-orbital model is consistent with the experimental observation that doped holes are located predominantly on oxygens, a result that goes beyond the one-orbital model. Nevertheless, the phase boundary between pseudogap and correlated metal, the Widom line, and the origin of the pairing energy (kinetic vs potential) are similar to the one-orbital model, demonstrating that these are emergent phenomena characteristic of  doped Mott insulators, independently of many microscopic details. Broader implications are discussed.
\end{abstract}
\maketitle


The appearance of high-temperature superconductivity upon doping an interaction-driven insulator is one of the most surprising phenomena in nature. A major goal of research in that field is to chart the phase diagram in the hope of providing key insights into an unconventional pairing mechanism and into the nature of the strongly correlated states of matter observed. 
With experiments driving this quest, and revealing a complex phase diagram~\cite{keimerRev}, theory is challenged to provide a framework to explain such complexity. The challenge comes from the fact that the insulating phase that is doped arises from interactions so strong~\cite{anderson:1987} that tools to describe such a non-perturbative regime are needed. Significant progress has been made in this area by novel theoretical approaches such as cluster extensions~\cite{maier,kotliarRMP,tremblayR} of dynamical mean-field theory (DMFT)~\cite{rmp}. 

The physics that must be understood is that of a square lattice made of CuO$_2$ unit cells where electrons on copper interact strongly. Intense effort devoted to study the case of a single orbital per unit-cell with an on-site repulsion, i.e. the two-dimensional Hubbard model, has shown that this simple model captures the basic phenomenology of cuprates~\cite{GullNewsViews:2015,AMJulich}. 

A more realistic model for the cuprates includes three orbitals per CuO$_2$ unit cell~\cite{Emery_1987, Varma_1987}. The necessity of this model is demonstrated  by numerous experiments that show that doped holes are found on oxygen~\cite{Gauquelin2014}. 
The ability to delocalize on oxygen allows electrons to feel a much weaker effective interaction, but at one-hole per unit-cell and strong-enough repulsion on copper, one obtains a charge-transfer insulator~\cite{zsa}. It is this kind of interaction-driven insulator of the three-orbital model that becomes a high-temperature superconductor upon doping. Hence this is the model we study.

Previous single-site DMFT calculations~\cite{AntoineCuO2, Lombardo1996, Zolf2000, sar, Cedric2008, Weber:2010, DeMedici2009, Wang:2011, Wang:2011b} provided important insights on the phase diagram. However the inclusion of short-range correlations is still a formidable theoretical problem. Despite pioneering investigations using cluster methods~\cite{Macridin2005, ArrigoniCuO2, Weber2011, go}, the precise form of the temperature-doping $T$-$\delta$ phase diagram is largely unexplored and several of its key aspects are uncertain. Notably, the finite temperature behavior of the metal to charge-transfer insulator transition driven by hole doping is unknown. 

Here we chart the cellular DMFT solution of the whole $T$-$\delta$ phase diagram of a doped charge-transfer insulator. We focus on four possible phases of the model, namely the charge-transfer insulator, the pseudogap, the correlated metal, and a d-wave superconducting state, along with their phase boundaries. Our goal is to establish if the basic phenomenology of cuprates found in the one-orbital model survives in the more realistic three-orbital model and which phenomena are emergent, i.e. independent of such details as number of orbitals per unit cell, shape of the Fermi surface, redistribution of spectral weight, location of holes within the unit cell.

%

\section{Model and method}
We consider the three-band Hamiltonian for copper $3d_{x^2-y^2}$ and oxygen $2p_x$,$2p_y$ orbitals. Ordering the corresponding annihilation operators as $(d_{\mathbf{k}\sigma}, p_{x,\mathbf{k}\sigma}, p_{y,\mathbf{k}\sigma})$, where $\mathbf{k}$ is the wave-vector and $\sigma$ the spin, the non-interacting part of the Hamiltonian for the infinite  lattice reads~\cite{AndersenLDA}:
\begin{equation}
\begin{aligned}
\mathbf{h}_0(\mathbf{k}) =   \left (
   \begin{array}{ccc}
    \epsilon_d & V_{dp_x} & V_{dp_y}  \\
    V_{dp_x}^\dag & \epsilon_p + W_{p_xp_x} & W_{p_xp_y} \\
    V_{dp_y}^\dag & W_{p_xp_y}^\dag & \epsilon_p + W_{p_yp_y}
   \end{array}
\right),
\label{h_0}
\end{aligned}
\end{equation}
with $V_{dp_x}=t_{pd}(1 - e^{ik_x})$, $V_{dp_y}=t_{pd}(1 - e^{ik_y})$, $W_{p_xp_x}=2t_{pp}(\cos k_x - 1)$, $W_{p_yp_y}=2t_{pp}(\cos k_y - 1)$ and $W_{p_xp_y}=t_{pp}(1 - e^{-ik_x})(1 - e^{ik_y})$. Here $t_{pd}$ ($t_{pp}$) is the oxygen-copper (oxygen-oxygen) hopping amplitude and $\epsilon_d$ ($\epsilon_p$) is the copper (oxygen) on-site energy. The copper-copper distance and $t_{pp}$ are taken as unity. 
$\mathbf{h}_0$ leads to the Fermi surface observed experimentally in the overdoped region of the cuprates (see supplementary Fig.2). 
For the interacting part, only the on-site repulsion on $d$ orbitals $U_{d}$ is retained. 

We solve this model with cellular dynamical mean-field theory (CDMFT), which isolates a cluster of 12 lattice sites with $(N_d, N_p) = (4, 8)$, and replaces the missing lattice environment by a self-consistent non-interacting bath. The cluster plus bath impurity model is solved with continuous-time quantum Monte Carlo for the hybridization expansion~\cite{millisRMP}. 
See supplementary Sec.I for details.


\section{Opening of the charge-transfer gap}
As described by the Zaanen-Sawatzky-Allen scheme~\cite{zsa} (ZSA), this model accounts for different correlated insulating states when the total occupation is $n_{\rm tot}=n_d+2n_{p}=5$ (one hole per CuO$_2$ unit): the charge-transfer insulator and the Mott-Hubbard one. The former is relevant for the cuprates and is the focus of the present work. 

\begin{figure}[t!]
\centering{
\includegraphics[width=0.98\linewidth,clip=]{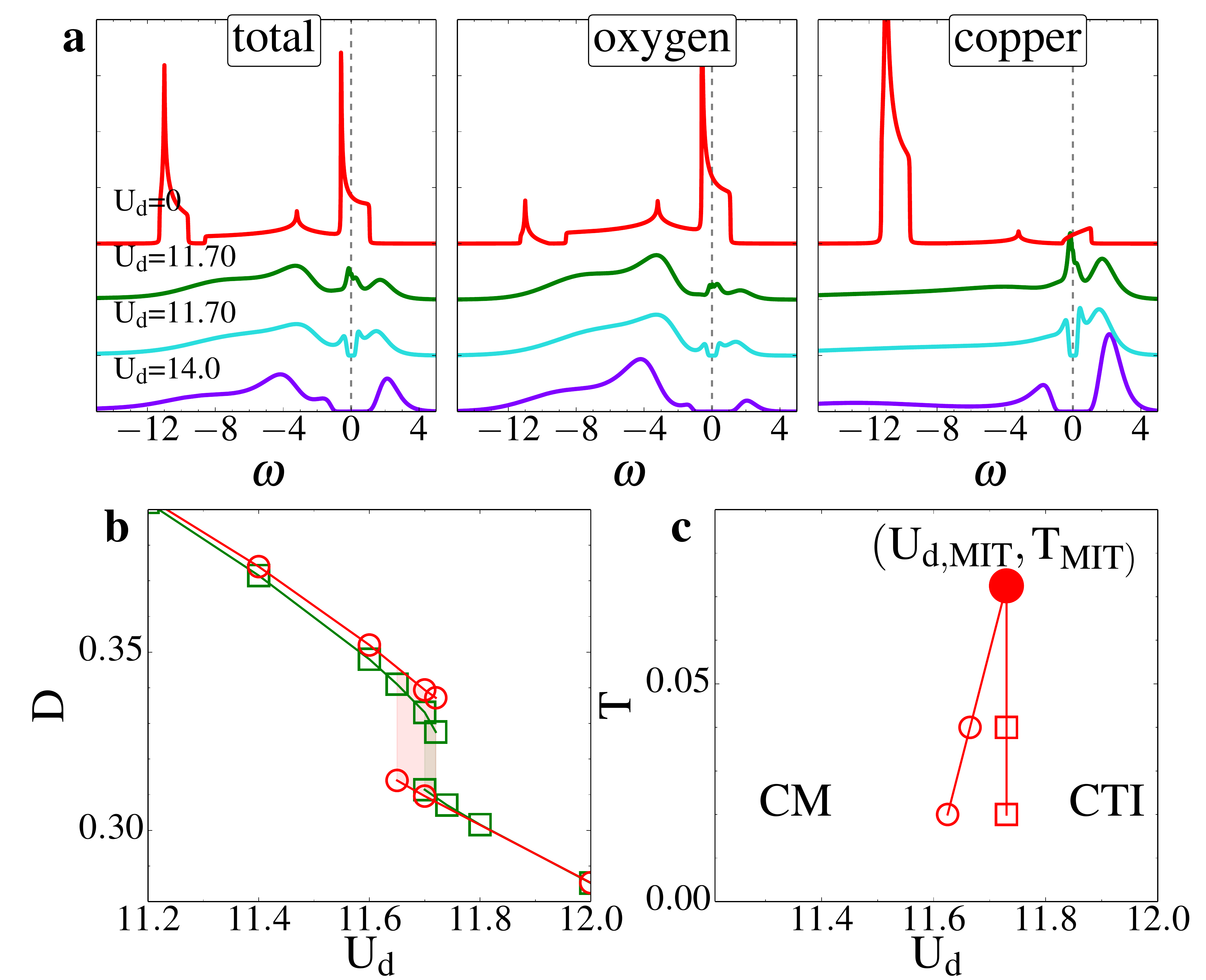}}
\caption{
(a) Local density of states $N(\omega)$ at $n_{\rm tot}=5$ and $\beta=50$ for several values of $U_d$. From left to right: total, projected $N(\omega)$ on the $p$ and $d$ orbitals. Other model parameters are $|\epsilon_p-\epsilon_d| =9$, $t_{pp}=1$ and $t_{pd}=1.5$. 
(b) Double occupancy $D$ as a function of $U_{d}$ at $n_{\rm tot}=5$ for $\beta=25$ (squares) and $\beta=50$ (circles). Hysteresis region is shaded. 
(c) $T$ versus $U_{d}$ phase diagram at $n_{\rm tot} =5$. A first-order transition at finite $U_d$ between a correlated metal (CM) and a charge-transfer insulator (CTI) is bounded by the jumps in the double occupancy and terminates at a critical endpoint. 
}
\label{fig1}
\end{figure}

Fig.\ref{fig1}a shows the local density of states (DOS) $N(\omega)=-1/\pi \rm{Im}G(\omega)$ at $n_{tot}=5$ for several values of $U_{d}$ at the inverse temperature $\beta=50$ (from left to right: total, projected DOS on the $p$ and $d$ orbitals).
The zero of energy is the Fermi level. 
To set the system in the charge-transfer regime, we take $\epsilon_d=0$ and  $\epsilon_p=9$ so that the localised $d$ orbital is beneath the oxygen band. The bandwidth originating from $t_{pp}$ alone is 8. By virtue of the hybridization term, here $t_{pd}=1.5$, the $d$ electrons acquire a finite dispersion and, for the noninteracting case, form a narrow band centered at $\omega \approx-11$ (see upper red curve and supplementary Sec.I). 
As described by the ZSA scheme~\cite{zsa}, for $U_{d} > |\epsilon_p -\epsilon_d|$ and $|\epsilon_p -\epsilon_d| > t_{pd}$ a correlation gap opens up and the system becomes a charge-transfer insulator. The lower violet curve shows this dramatic effect of correlations. 
At low temperature, the interaction-driven transition between the metal and the charge-transfer insulator is first-order. At intermediate values of $U_{d}$, there is a coexistence between a metallic and an insulating solutions to the CDMFT equations (green and blue curve, respectively). 

%

The first-order nature of the transition is best shown by the double occupancy $D$ of $d$ orbitals as a function of $U_{d}$ (cf Fig.\ref{fig1}b). 
$D$ shows hysteresis loops between two solutions, with sudden jumps at $U_{d, c1}(T)$, where the insulating solution ceases to exists, and at $U_{d, c2}(T)$, where the metallic solution disappears. Hysteresis loops become wider with decreasing $T$.
The behavior of $D$ allows us to construct the temperature versus $U_{d}$ phase diagram in Fig.\ref{fig1}c. The first-order transition between a correlated metal and a charge-transfer insulator occurs within the coexistence region, bounded by the spinodals $U_{d,c1}$ and $U_{d,c2}$ (open circles and squares, respectively), and terminates in a critical endpoint, where $dD/dU_d$ diverges. Our results extend to finite temperature the previously obtained $T=0$ phase diagram~\cite{go}.

\begin{figure}[ht!]
\centering{
\includegraphics[width=0.98\linewidth,clip=]{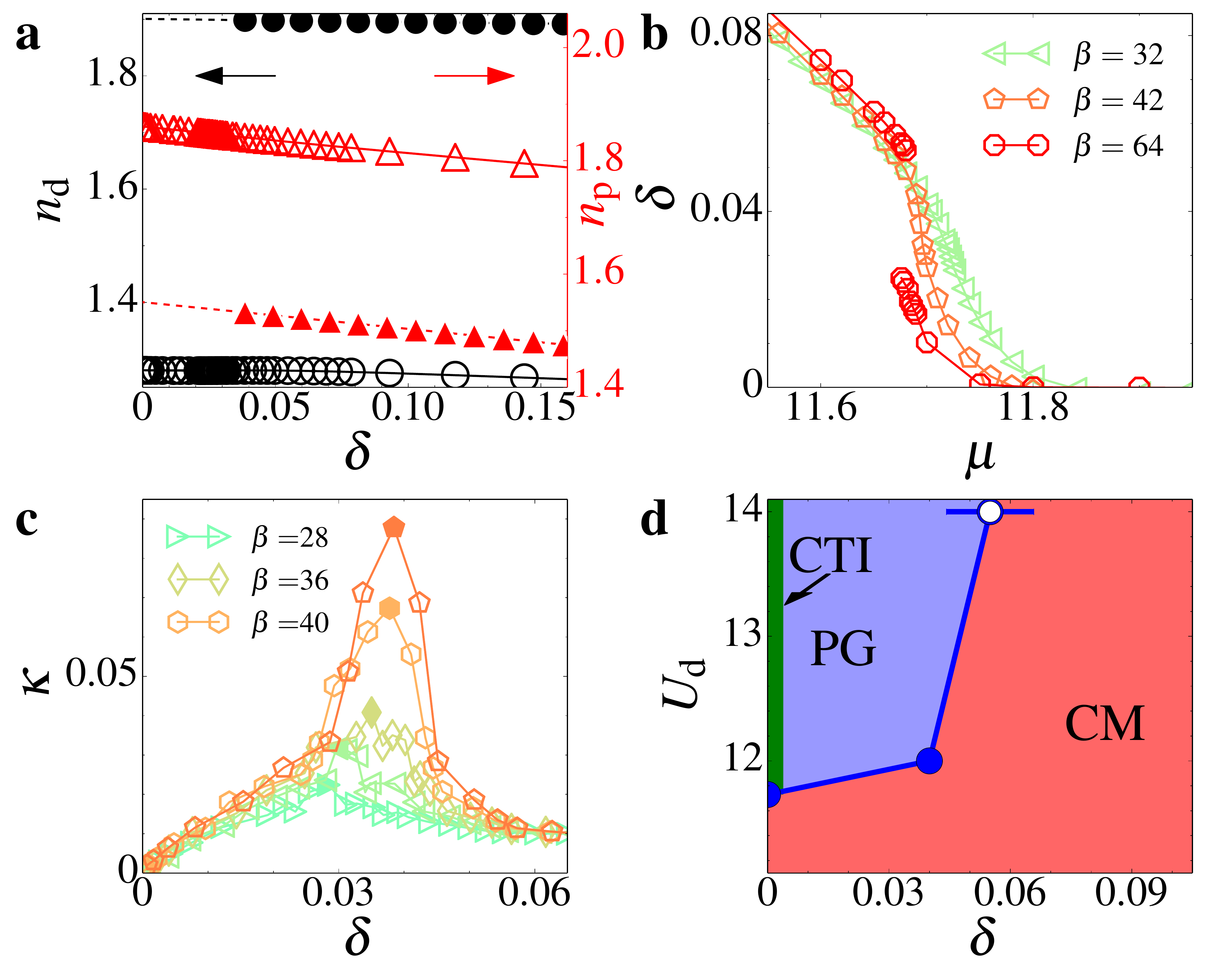}}
\caption{
(a) Partial occupation $n_{d}$ (circles), $n_{p}$ (triangles) versus $\delta=5-n_{\rm tot}$ at $\beta=25$ and $U_{d}=0, 12$ (full and open symbols, respectively). 
(b) $\delta$ versus $\mu$ for $U_{d}=12$, for different temperatures. A plateau at $\delta(\mu)=0$ signals the CTI. Hysteresis appears at finite doping at the lowest $T$. 
(c) Charge compressibility $\kappa$ versus $\delta$ for different temperatures at $U_{d}=12$. $\kappa$ diverges at the endpoint $(\delta_p,T_p)$ of the PG-CM first-order transition. 
(d) $U_{d}$ versus $\delta$ phase diagram. The boundary between CTI and PG is second-order. The boundary between CTI and CM at $\delta=0$ is first-order. Green line is drawn at $T=T_{\rm MIT}. $The boundary between PG and CM is first-order. Blue line denotes $\delta_p$ versus $U_d$. Since $T_p$ decreases with increasing $U_d$, we show as blue solid circles the position $\delta_p$ of the endpoints and as blue open circle the position of $\kappa_{\rm max}$ at our lowest $T$. The full data set for (b) and (c) is in supplementary Fig.3 and Fig.4 respectively. Model parameters are $|\epsilon_p-\epsilon_d| =9$, $t_{pp}=1$ and $t_{pd}=1.5$.
}
\label{fig2}
\end{figure}

\section{Hole-doping driven metal-insulator transition}
Fig.\ref{fig2}a shows the partial occupation of oxygen and copper as a function of hole doping $\delta= 5-n_{\rm tot}$ for $U_d=12$ and $\beta=25$ (open symbols). 
In the undoped system, comparison with $U_d=0$ (full symbols) shows that at finite $U_d$ electrons are transfered from copper to oxygen. Lowering the chemical potential $\mu$ results in an essentially doping-independent $n_{d}$, while $n_{p}$ decreases, indicating that the holes mainly enter the oxygen, as expected in the charge-transfer regime and found experimentally~\cite{Gauquelin2014}. 

Fig.\ref{fig2}b shows the doping  as a function of $\mu$ for $U_{d}=12$ and different temperatures. 
The plateau in the curves at $\delta(\mu)=0$ reveals the onset of the incompressible charge-transfer insulator. By lowering $\mu$ until we obtain a compressible state, the isotherms $\delta(\mu)$ evolve continuously, i.e. without hysteresis. Hence we conclude that the transition between charge-transfer insulator and compressible phase is of second-order at $T=0$. The latter has the characteristics of a pseudogap phase, as discussed below. 
Upon doping further, a first-order transition occurs between that pseudogap phase and a more conventional correlated metal. Indeed, as $T$ decreases, the isotherms $\delta(\mu)$ develop a sigmoidal shape and eventually hysteresis between two compressible solutions. 
This transition ends in a second-order critical point at $(\delta_p,T_p)$, at which thermodynamic response functions, such as the charge compressibility $\kappa = 1/n_{tot}^2 (dn_{tot}/d\mu)_T$, shown in Fig.\ref{fig2}c versus $\delta$, diverge. 
For $T>T_p$, the two phases merge in a single supercritical phase, and the divergence in $\kappa$ is replaced by a maximum value, which decreases with increasing $T$.  
It is striking that either below or above $T_p$,  $(d\delta_{tot}/dT)_\mu$ changes sign, from positive at small doping to negative at large doping, as can be seen from Fig.\ref{fig2}b. 

The emergent first-order transition at finite doping is connected to the charge-transfer insulator to metal transition in the undoped case. This can be deduced by tracking the position $\delta_p$ of the critical endpoint as a function of $U_d$, as shown by blue circles in Fig.\ref{fig2}d. 
This line of critical endpoints (where the pseudogap-correlated metal transition ends) starts out at the metal to charge-transfer insulator transition at $\delta=0$ and $U_{d}\simeq 11.6$ and moves progressively to high doping and lower $T$ as $U_{d}$ increases.

\begin{figure}
\centering{
\includegraphics[width=0.98\linewidth,clip=]{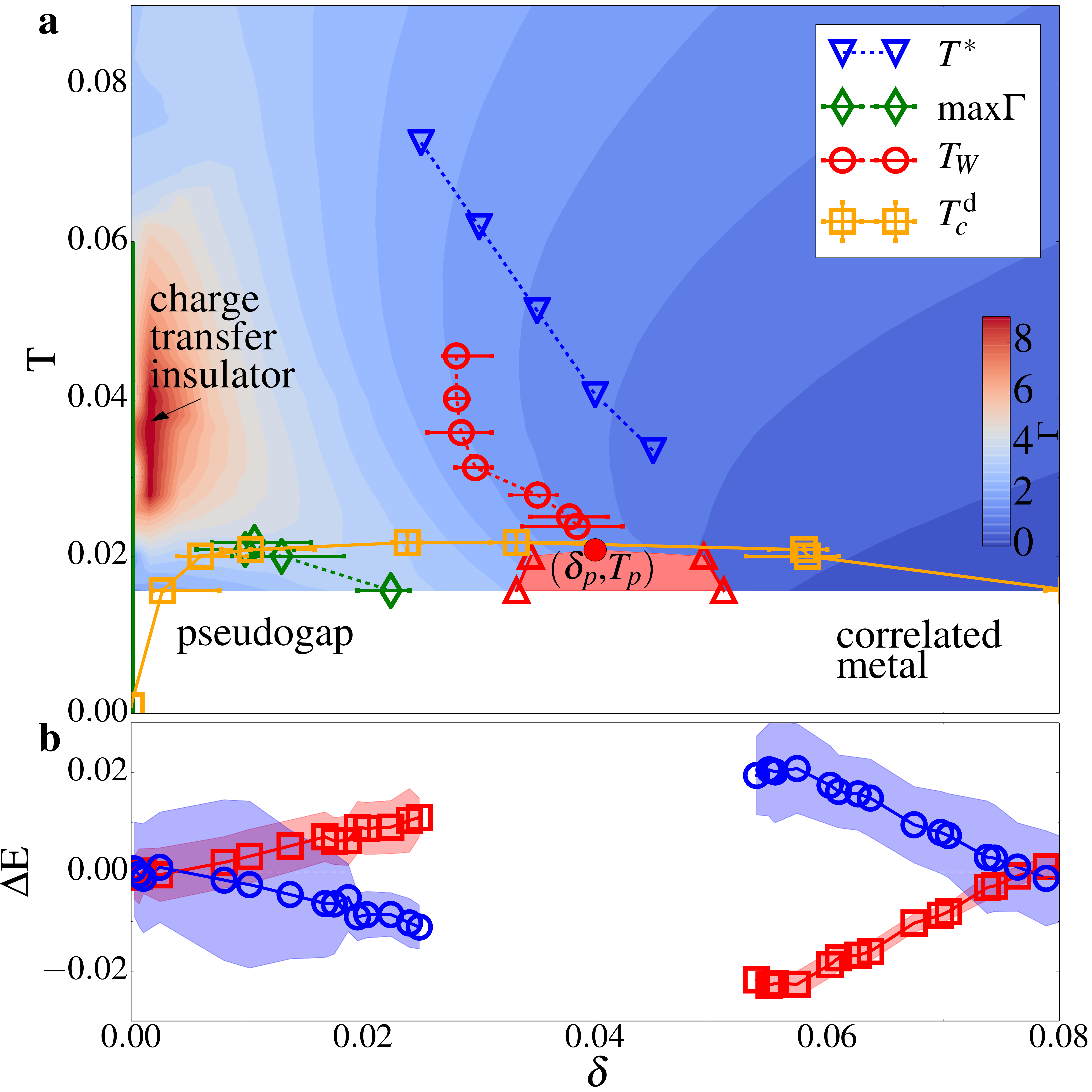}}
\caption{
(a) Temperature versus hole doping phase diagram. Model parameters are $|\epsilon_p-\epsilon_d| =9$, $t_{pp}=1$, $t_{pd}=1.5$ and $U_d=12$. 
Four phases are shown. The d-wave superconducting phase is determined by a non-zero superconducting order parameter, and is delimited by $T_c^d$ (orange squares). The three normal-state phases are determined by the behavior of $\delta(\mu)$: CTI at $\delta=0$, PG and CM. Below $T_c^d$ the normal-state is metastable. PG and CM are separated by a first-order transition at finite doping (red triangles denote the jump in the occupation curves), terminating at a critical endpoint $(\delta_p,T_p)$ (full circle). Emanating from it is $T_W$, the crossover line of the maxima of the charge compressibility $\kappa$ (open red circles), which is a proxy for the Widom line. Its high temperature precursor is $T^*$, the line where the DOS at the Fermi level drops as a function of $T$ (blue triangles). Color corresponds to the magnitude of the normal-state scattering rate $\Gamma$ at cluster momentum $K=(\pi,0)$.  Green diamonds indicate the maximum of $\Gamma(\delta)|_T$ at low $T$ and $\delta>0$.
(b) Difference in kinetic and potential energies between the superconducting and normal states (blue and red lines, respectively) versus $\delta$ at $\beta=64$. Shaded regions give standard errors.
}
\label{fig3}
\end{figure}
\begin{figure*}[!ht]
\centering{
\includegraphics[width=0.98\linewidth,clip=]{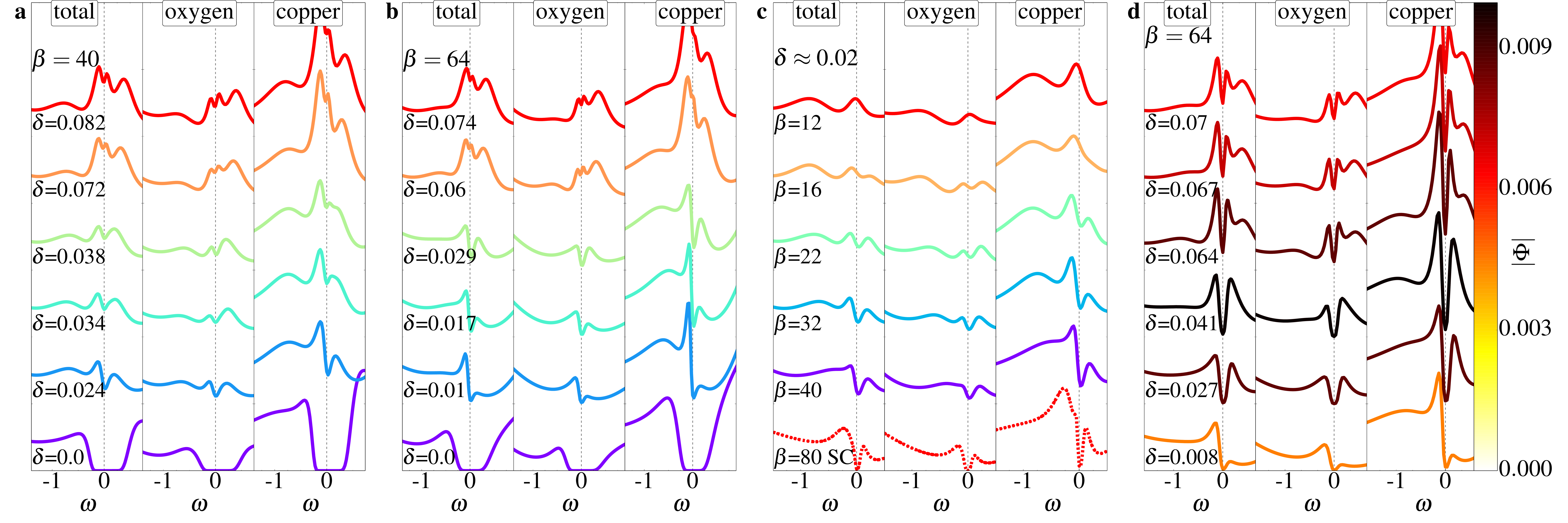}}
\caption{Low frequency part of the local DOS $N(\omega)$. Each DOS is normalized to unity. $N(\omega)_{tot}=\frac{2}{3}N(\omega)_{p}+\frac{1}{3}N(\omega)_{d}$.
(a,b) $N(\omega)$ for different dopings at constant inverse temperature (a) $\beta=40<1/T_p$ and (b) $\beta=64>1/T_p$. 
(c) $N(\omega)$ for different temperatures at constant doping $\delta \approx 0.02$.
(d) $N(\omega)$ in the superconducting state at $\beta=64$ for different dopings. In this panel, color corresponds to the magnitude of the superconducting order parameter. 
}
\label{fig4}
\end{figure*}

\section{Phase diagram}
The temperature-doping phase diagram shown in Fig.\ref{fig3}a summarizes the normal-state properties investigated so far. 
At zero doping, a second-order transition separates a charge-transfer insulating phase from a pseudogap phase. 
At finite doping, there is a first-order transition between two normal-state phases: the pseudogap and the correlated metallic state (lines with triangles are an estimate for the spinodal boundaries determined from the jumps in $\delta(\mu)$). 
As the temperature increases, the first-order transition ends in a second-order critical point at $(T_p,\delta_p)$, where the thermodynamic response functions, such as the charge compressibility $\kappa$ discussed above, diverge. 
For $T>T_p$, only one supercritical phase exists, but the first-order transition generates a crossover, the Widom line $T_W$, at which thermodynamic response functions show maxima (red line with circles is the line of maximum of $\kappa$ computed in Fig.\ref{fig2}c). 
Quite generally, $T_W$ marks also the border between different dynamic behaviors~\cite{water1,ssht}: the drop in the local DOS goes through an inflection point at $T_W$ (supplementary Fig.7). The onset of the drop in such quantity commonly defines the onset of the pseudogap $T^*$ and occurs at the higher precursory temperature (see line with blue triangles and supplementary Fig.7c). 
This is qualitatively consistent with experiments~\cite{Alloul2013}. 
The development of the pseudogap is characterised by the growth of inter-site self-energies or, equivalently, by a strong momentum space differentiation of the electronic lifetimes (see supplementary Figs.5, 6). 
The first-order transition at finite doping between pseudogap and correlated metal, and its associated crossover, is the unifying feature of self-energy anisotropy, as can be seen by the ridge of large scattering rate $\Gamma_{K=(\pi,0)}$, shown as color plot in Fig.\ref{fig3}a, emerging from $\delta_{c1}(T\rightarrow 0)$ and bent toward the charge-transfer insulator (line with green diamonds).  

Proximity to Mott physics often entwines with broken symmetry states. While speculating that along the charge-compressibility maxima $T_W$, the charge should be most susceptible to develop charge-density modulations, we restrict our study to d-wave superconductivity. 
The dynamical mean-field superconducting transition temperature $T_c^d$ (orange line in Fig.\ref{fig3}a) marks the temperature below which the superconducting order parameter is nonzero (see supplementary Fig.13) and corresponds to the temperature below which Cooper pairs develop within the cluster. 
Complex behavior in the superconducting state originates from the underlying normal-state first-oder transition: 
(a) $T_c^d$  forms an asymmetric dome as a function of doping, whose broad maximum occurs close to the intercept of maximum $\Gamma$ (green line). 
(b) $T_W$  intercepts $T_c^d$, indicating that superconductivity and pseudogap are distinct phenomena, although they are entwined ones, since in our approach the origin of both phenomena is rooted in Mott physics. 
(c) For $\delta<\delta_p$, pairing is driven by kinetic energy, while for $\delta>\delta_p$ it is driven by potential energy, as illustrated in Fig.\ref{fig3}b by  the difference of potential and kinetic energy between superconducting and normal states as a function of doping. This is qualitatively consistent with optical measurements~\cite{Molegraaf2002, giannetti2011}.

The local DOS, shown in Fig.\ref{fig4} (for the entire frequency spectrum, see supplementary Figs.~9-12) helps define the various phases. 
Figs.\ref{fig4}a,b show the doping evolution of the normal-state DOS above and below $T_p$, respectively. Note that 
(i) At $\delta=0$, $N(\omega)$ shows a correlation gap of charge-transfer type; 
(ii) Upon hole doping, a dramatic redistribution of spectral weight occurs in $N(\omega)$, as a consequence of electronic correlations. At low frequency $N(\omega)$ develops a pseudogap 
having a large Cu character with significant O component~\cite{ZhangRice1988, Macridin2005}. The frequency profile has a large particle-hole asymmetry, qualitatively similar to experimental observation~\cite{Davis:2007}. 
(iii) Increasing $\delta$ further, particle-hole asymmetry is reduced, the spectral weight inside the pseudogap progressively fills in, eventually the pseudogap disappears and a broad peak at the Fermi level takes shape. 

Fig.\ref{fig4}c shows the temperature evolution of $N(\omega)$ for $\delta<\delta_p$ and demonstrates that the pseudogap gradually fills in upon raising $T$, in qualitative agreement with experiment~\cite{timusk}. Last, Fig.\ref{fig4}d shows $N(\omega)$ in the superconducting state for several dopings. The coherence peaks are visible both in Cu and O partial DOS, demonstrating that Cooper pairs are composite objects of mixed d-p character. This feature is reminiscent of ``Zhang-Rice singlet'' physics~\cite{ZhangRice1988}. 

The present study shows that an antiferromagnetic quantum critical point~\cite{Chubukov2008} is not necessary to obtain pseudogap or d-wave superconductivity. The interplay with other broken-symmetry phases, such as charge density waves~\cite{efetov_pseudogap_2013, SachedevBond2013, ChubukovCDW2014}, or loop currents~\cite{Varma1989, Varma2006, Cedric2014, Bulut2014, Kung2014, Pepin2016} are important issues to be considered in subsequent investigations.


\section{Summary}
In summary, we charted the phase diagram of a hole-doped charge-transfer insulator using a three-band model solved with CDMFT. 
We revealed the structure of both the normal and superconducting phases, and fingerprinted their organizing principle as a normal-state first-order transition below the superconducting dome. 
This transition is analog to the one found in the single band Hubbard model~\cite{sht, sht2, ssht, sshtSC, sshtRHO, LorenzoSC}, despite the large differences in microscopic details, namely the presence of oxygen, the different band structure, and the energy redistribution of spectral weight. 
This suggests the emergent character of the phenomenon, solely produced by Mott physics plus short-range correlations, and thus leads to the following conjecture:
A first-order transition, even when hidden by another phase, here superconductivity, can act as a general organizing principle of strongly-coupled matter. In solids this is now clear for systems described by Hubbard-like models~\cite{ssht,LorenzoSC,furukawaWL,vlad1,vlad2,vlad3,Hebert:2015},  in fluids it has been suggested before~\cite{water1,supercritical,Simeoni2010}, and in quark matter the putative critical endpoint in the $T$ vs baryon chemical potential that appears in the QCD phase diagram may be another manifestation of the generality of this phenomenon in strongly interacting systems~\cite{Lacey2015, Stephanov1998, StephanovRev}.

\begin{acknowledgments}
We acknowledge D. S\'en\'echal, G. Kotliar, M. Rozenberg and Y. Sidis for discussions. This work was partially supported by Natural Sciences and Engineering Research Council (Canada), the Tier I Canada Research Chair Program (A.-M.S.T.) and Universit\'e de Sherbrooke. Simulations were performed on computers provided by CFI, MELS, Calcul Qu\'ebec and Compute Canada.
\end{acknowledgments}




%

 


\onecolumngrid
\clearpage
\setcounter{figure}{0}
\setcounter{section}{0}

\begin{center}

{\bf Supplementary information} 

{\bf Pseudogap and superconductivity in two-dimensional doped charge-transfer insulators}

{L. Fratino, P. S\'emon, G. Sordi, A.-M.S. Tremblay}

\end{center}

\vspace{0.1cm}
	
Section~\ref{SectionA} shows some details of the model and method. Section~\ref{SectionB} expands on the doping-driven metal to insulator transition. In Section~\ref{SectionC} we present the full frequency spectrum of the local DOS along with the superconducting order parameter. 

\section{Model and method}
\label{SectionA}

\begin{figure}[!h]
\centering{
\includegraphics[width=0.54\linewidth,clip=]{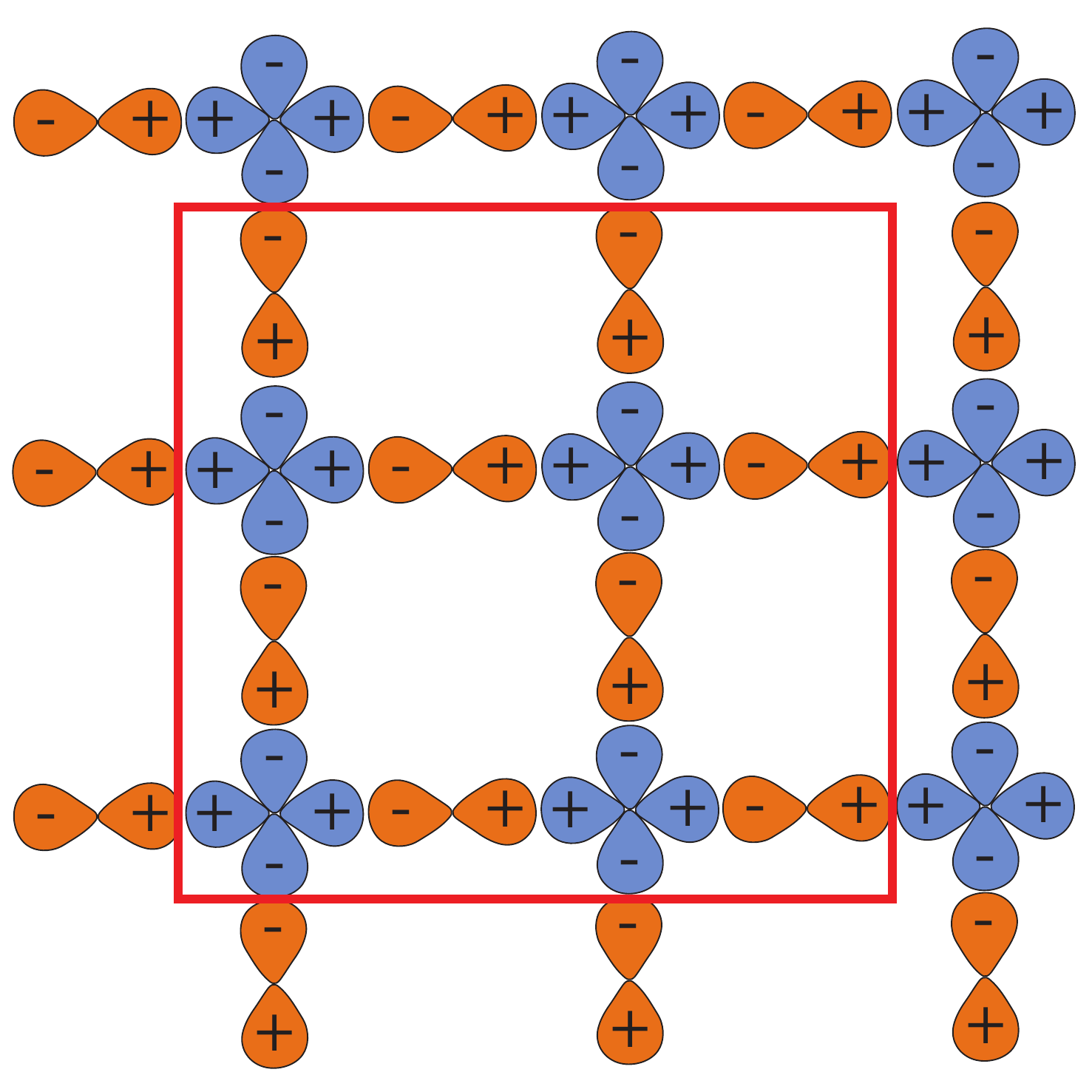}}
\caption{Sketch of the three-band model. Cu $3d_{x^2-y^2}$ orbitals and O $2p_x$, $2p_y$ orbitals are drawn in blue and orange, respectively. Our phase convention is indicated.  CDMFT takes a cluster of 12 sites with $(N_d, N_p) = (4, 8)$ (see red square) out of the lattice and embeds it in a self-consistent noninteracting bath. 
}
\label{figS1}
\end{figure}

We study the three-band Hamiltonian, (Fig.~\ref{figS1})
\begin{equation}
H =  \mathbf{h}_0 + U_d \sum_i n_{di\uparrow} n_{di\downarrow}
\end{equation}
where $\mathbf{h}_0$ is given by Eq.~(1) of main text, and the electron correlations are introduced by the local Coulomb repulsion $U_d$ on Cu sites. 
Note that in doing the Fourier transform to obtain Eq.~(1) of main text, we used the same phase for all atoms within a given unit cell. 

We solve this problem with cellular dynamical mean field theory (CDMFT)~\cite{maier_SM,kotliarRMP_SM,tremblayR_SM}, which isolates a cluster of 12 lattice sites with $(N_d, N_p) = (4, 8)$, and replaces the missing lattice environment by a non-interacting bath. 
Since there is no interaction on the oxygens, it is convenient to integrate them out before applying CDMFT. This yields an effective lattice action involving only copper $d$-orbitals. The Green function corresponding to the non-interacting part of this action is the uppermost diagonal element of the matrix Green function for the Hamiltonian Eq.~(1) of main text, namely  $G_{0\text{eff}}(i\omega_n,\mathbf{k}) = ((i\omega_n + \mu - \mathbf{h}_0(\mathbf{k}))^{-1})_{dd}$, with $i\omega_n$ Matsubara frequencies and $\mu$ the chemical potential. The action of the corresponding impurity model for CDMFT then reads
\begin{equation}
S_{\text{Imp}}= -\sum_{n\mathbf{R}\mathbf{R}'\sigma}d_{\mathbf{R}\sigma}^\dagger(i\omega_n)\mathcal{G}_{0,\mathbf{R}\mathbf{R}'}^{-1}(i\omega_n)d_{\mathbf{R}'\sigma}(i\omega_n) + U_{d} \sum_{\mathbf{R}} \int_0^\beta d^\dagger_{\mathbf{R}\uparrow}(\tau)d_{\mathbf{R}\uparrow}(\tau) d^\dagger_{\mathbf{R}\downarrow}(\tau)d_{\mathbf{R}\downarrow}(\tau)d\tau,
\label{equ:impurity}
\end{equation}
where $\mathbf{R}$ runs over the copper sites of the plaquette. 
The Weiss field $\mathbfcal{G}_{0}^{-1}$ is determined by the CDMFT self-consistency
\begin{equation}
(\mathbfcal{G}_{0}^{-1}(i\omega_n) - \mathbf{\Sigma}(i\omega_n))^{-1} =  \frac{N_d}{4\pi^2}\int (\mathbf{G}_{0\text{eff}}^{-1}(i\omega_n,\tilde{\mathbf{k}}) - \mathbf{\Sigma}(i\omega_n))^{-1}d\tilde{\mathbf{k}},
\end{equation}
where $\mathbf{\Sigma}$ is the self-energy of the impurity model and 
\begin{equation}
G_{0\text{eff},\mathbf{R}\mathbf{R}'}^{-1}(i\omega_n,\tilde{\mathbf{k}}) =  \frac{1}{N_d} \sum_{\mathbf{K}}e^{i(\mathbf{K} + \tilde{\mathbf{k}})(\mathbf{R} - \mathbf{R'})}G_{0\text{eff}}^{-1}(i\omega_n, \mathbf{K} + \tilde{\mathbf{k}})
\end{equation}
is the mixed real/momentum-space representation for the effective non-interacting lattice Green function $G_{0\text{eff}}(i\omega_n,\mathbf{k})$ defined above. The $\mathbf{K}$'s run over the momenta of the plaquette and the $\tilde{\mathbf{k}}$'s over the reduced Brillouin zone defined by the tiling of the lattice into plaquettes. Notice that the self-energy of the three-band model considered here is finite on the copper orbitals only and the interacting Green function (and filling) on oxygen orbitals can be obtained from the lattice Dyson equation involving all orbitals.

The impurity model Eq.~\eqref{equ:impurity} is solved with continuous-time  quantum Monte Carlo for the hybridization expansion\cite{millisRMP_SM}, where the Weiss field is written as  
%
$\mathbfcal{G}_{0}^{-1}(i\omega_n) = i\omega + \mu -\mathbf{t}_{cl} - \mathbf{\Delta}(i\omega_n)$,
%
such that the hybridization function $\mathbf{\Delta}(i\omega_n)\rightarrow 0$ for $n\rightarrow \infty$. Here the effective cluster hopping matrix $\mathbf{t}_{cl}$ is diagonal, as can be checked from the high-frequency limit of the self-consistency equation. Hence, the efficient segment picture applies for the evaluation of the trace over the cluster. This remains true when we allow for $d_{x^2-y^2}$ superconductivity by introducing the Nambu representation and anomalous hybridization functions that acquire non-zero components between nearest-neighbor copper sites. We measure the superconducting order parameter $\Phi=\pm\langle d_i^\dagger d_j^\dagger\rangle$ where i and j are nearest-neighbor copper sites and where the sign changes upon $\pi/2$ rotation. The symmetry is not broken inside the cluster.  

Physically, band-structure calculations of Ref.~\onlinecite{AndersenLDA_SM} and previous DMFT works (see Refs.13-25 of main text) suggest the following range of parameters in cuprates: $t_{pp} \approx 1.1$ eV, $t_{pd} \in (1.3-1.6)$ eV, $|\epsilon_p -\epsilon_d | \in  (3-5)$ eV, $U_d \in (8-9)$ eV. Our choice of parameters ($t_{pp}=1$, $t_{pd}=1.5$) is compatible with those values for $t_{pp}$ and $t_{pd}$. However we have chosen a larger value for the charge-transfer energy $|\epsilon_p - \epsilon_d | = 9$, and, as a consequence, a larger value of $U_d$ to open a charge-transfer gap. This is because we first want to focus on a clear ``charge-transfer insulator'' regime in the Zaanen-Sawatzky-Allen  framework~\cite{zsa_SM}, staying away from the ``intermediate region''. In fact, with our specific choice of parameters, in the limit $U_d=0$ and $t_{pd}=0$, the $d$ level lies just below the oxygen bands. A finite $t_{pd}$ then turns the $d$ level into a band with mainly $d$ character, and the conduction band still keeps mostly a $p$ character (see supplementary Fig.2 and red curves in Fig.~1a of main text). A smaller value of the charge-transfer energy $|\epsilon_p - \epsilon_d |$ will increase the mixed orbital character at the Fermi level.

\vspace{0.4cm}
\begin{figure}[!hb]
\centering{
\includegraphics[width=0.999\linewidth,clip=]{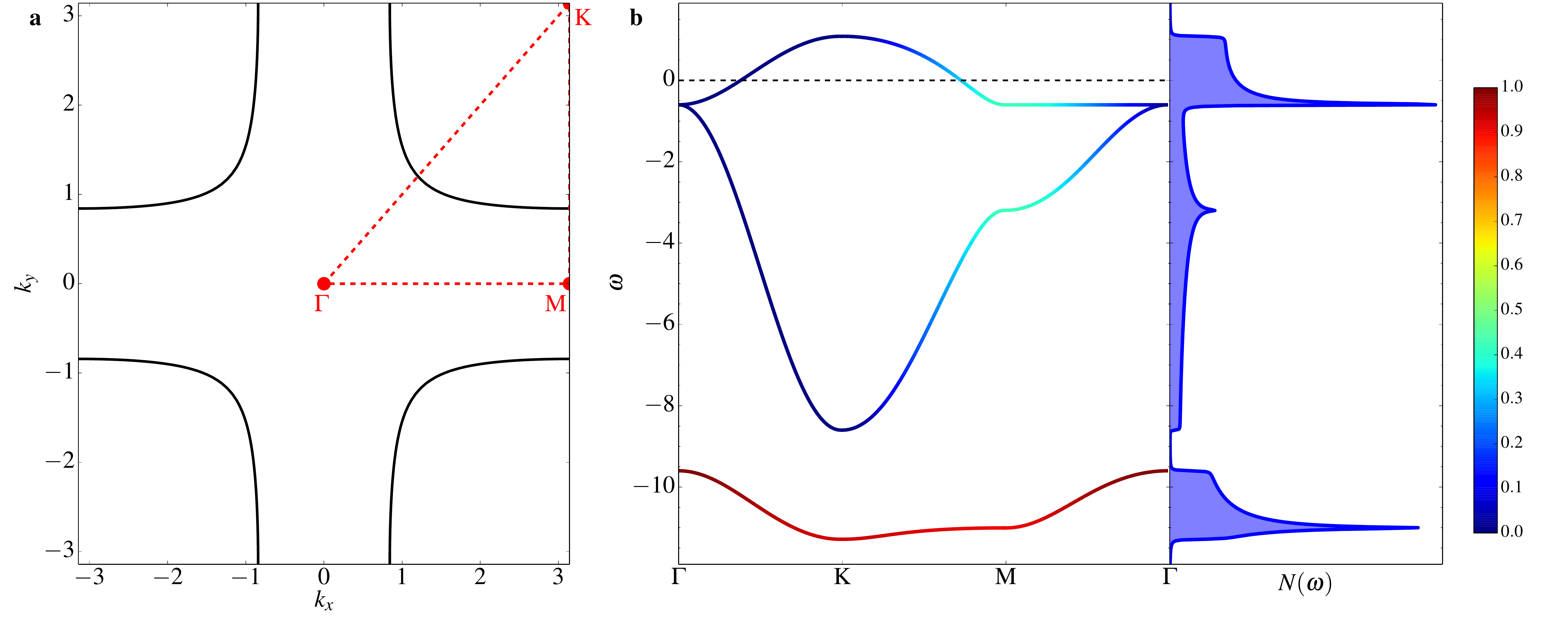}}
\caption{(a) Noninteracting Fermi surface for the model parameter investigated in Fig.~1a of main text, namely $\epsilon_p=9$, $t_{pp}=1$, $t_{pd}=1.5$, which gives a total occupation $n_{tot}$ equal to five. 
(b) Non-interacting band structure for the same model parameter along with the resulting total density of states. Color corresponds to the d-character of the hybridised bands. The band crossing the Fermi level has mostly oxygen character. 
}
\label{figS2}
\end{figure}
%



\section{Doping-driven metal-insulator transition}
\label{SectionB}
\begin{figure}[!h]
\centering{
\includegraphics[width=0.69\linewidth,clip=]{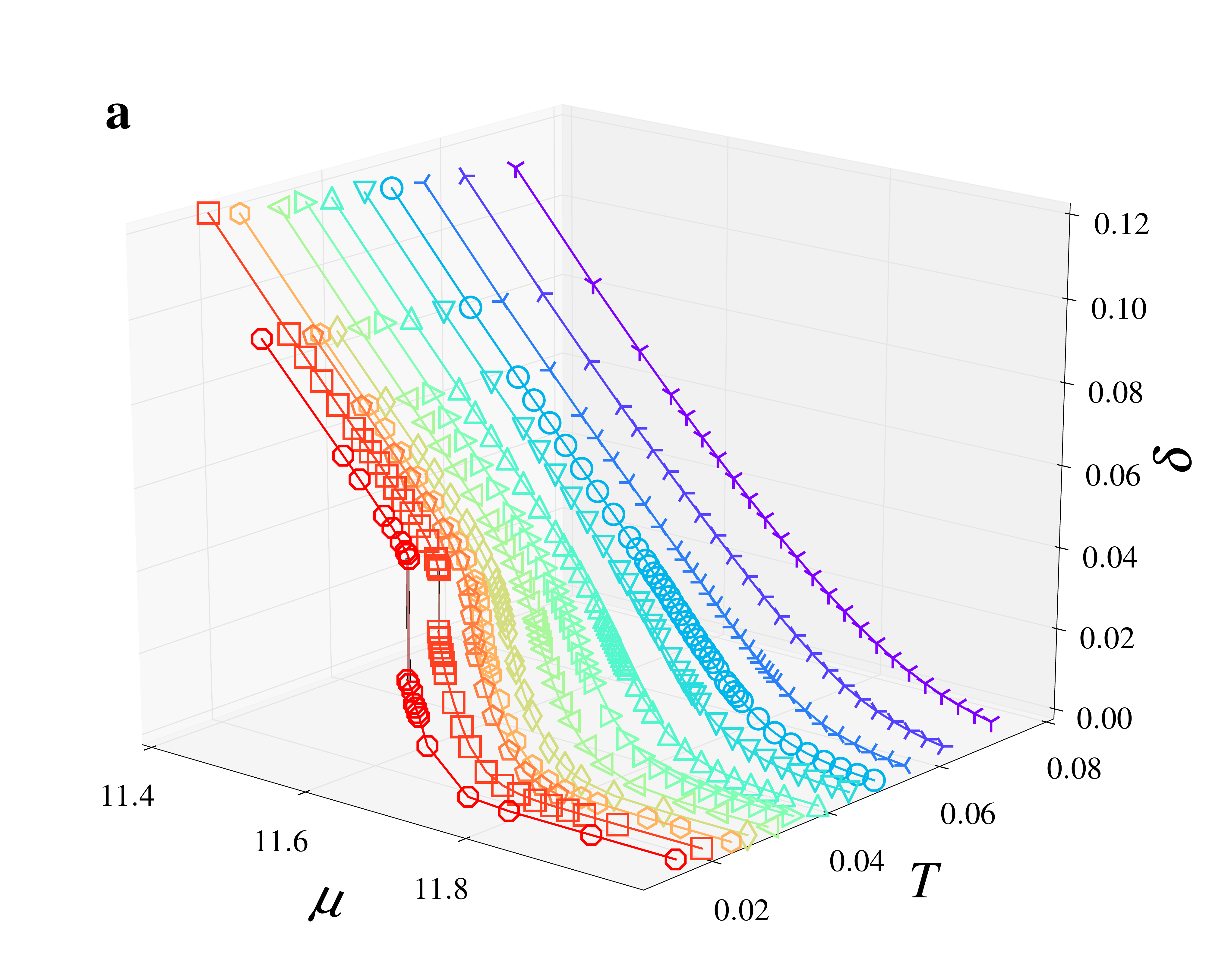}
\includegraphics[width=0.30\linewidth,clip=]{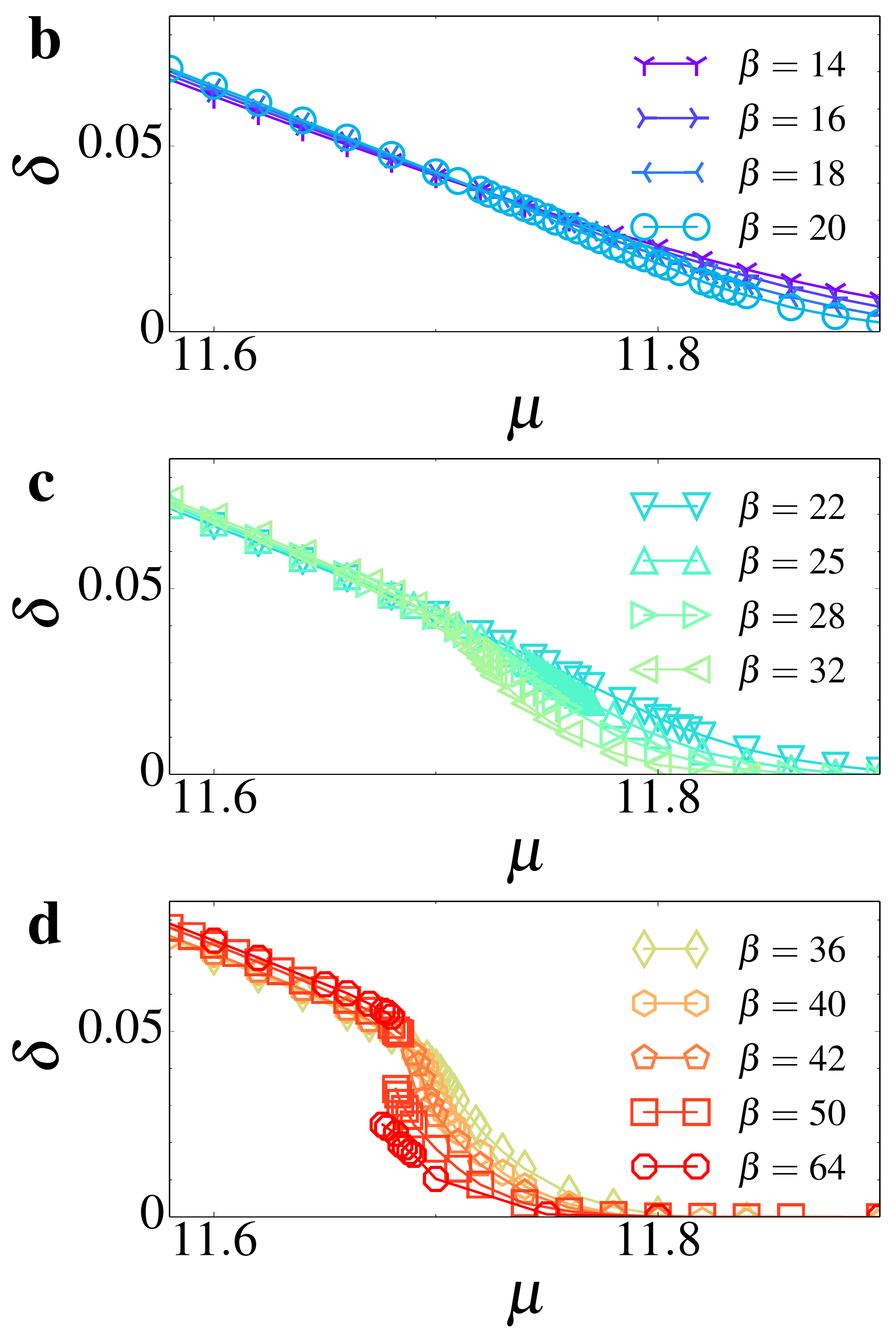}}
\caption{Extended data of figure~2b of main text. Left: doping versus $\mu$ for several temperatures $T$. Right: 2D projections.}
\label{figS3}
\end{figure}
\begin{figure}[!h]
\centering{
\includegraphics[width=0.69\linewidth,clip=]{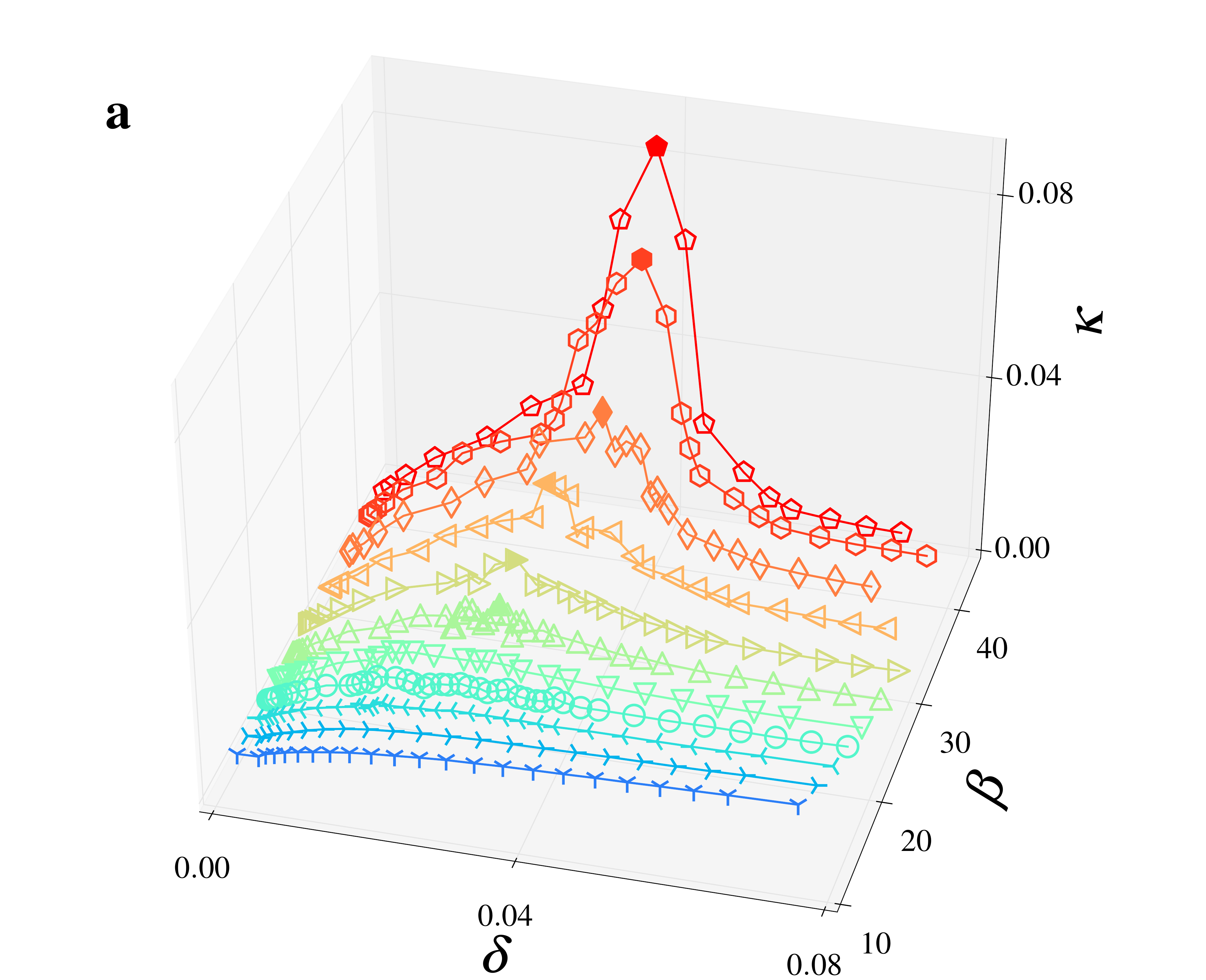}
\includegraphics[width=0.30\linewidth,clip=]{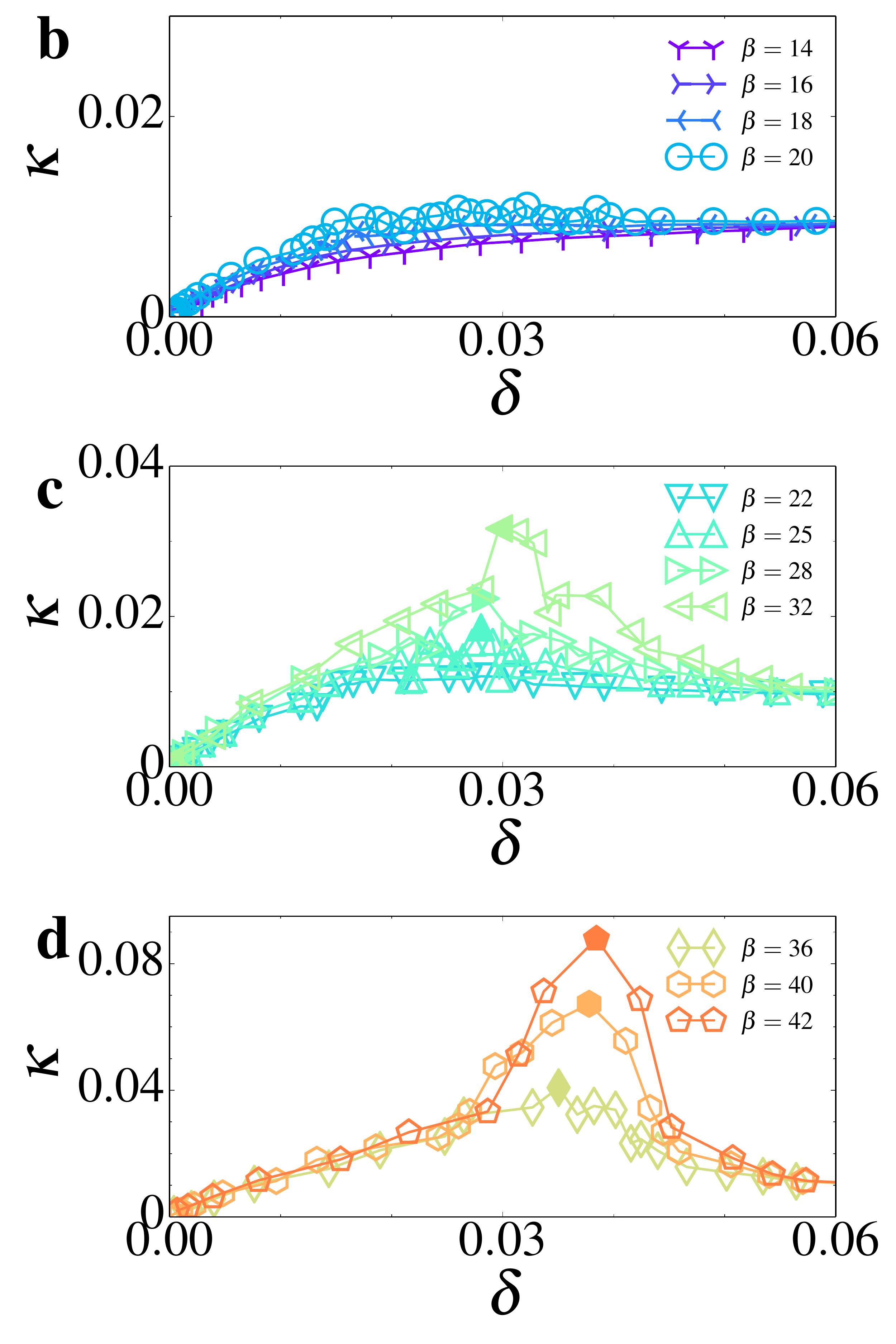}}
\caption{Extended data of figure~2c of main text. Left: charge compressibility $\kappa$ versus $\delta$ for several inverse temperatures $\beta$. Right: 2D projections.}
\label{figS4}
\end{figure}
\begin{figure*}[!h]
\centering{
\includegraphics[width=0.999\linewidth,clip=]{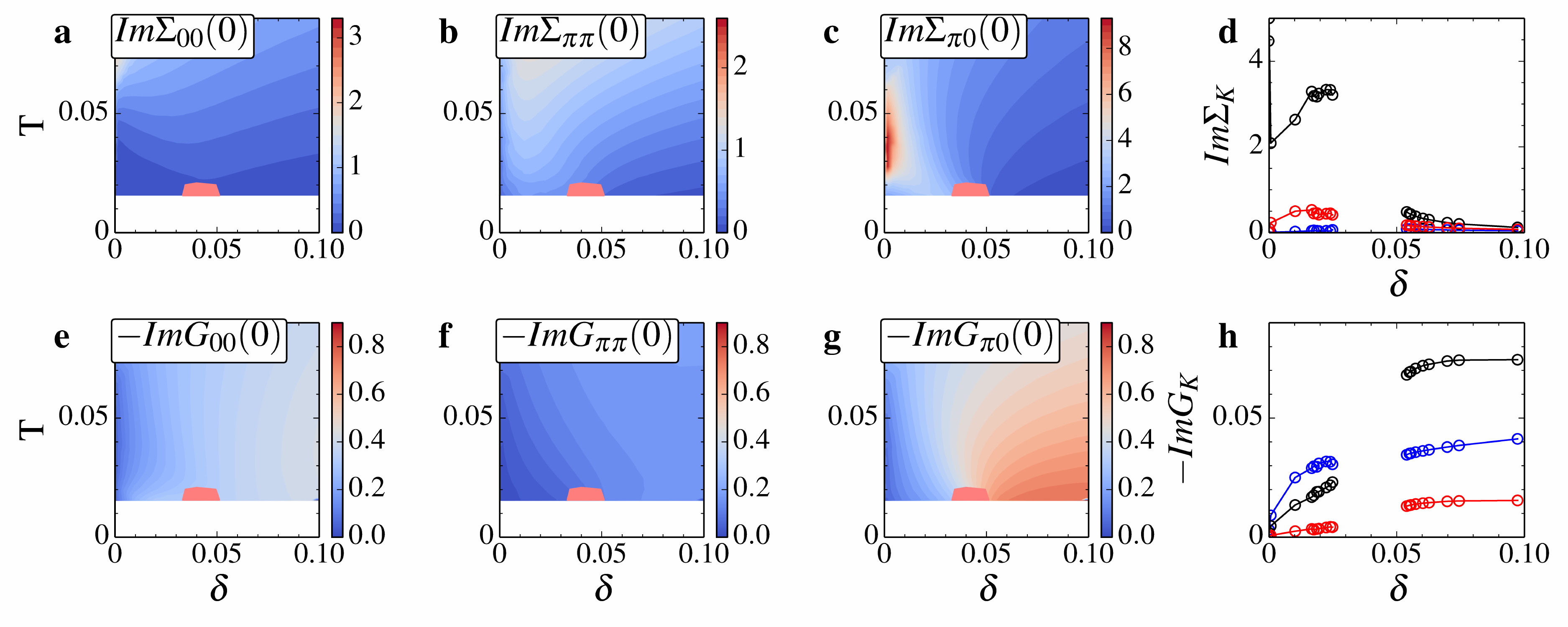}}
\caption{(a), (b), (c): Temperature versus hole doping colormap of the extrapolated zero-frequency value of the imaginary part of the cluster self-energy -Im$\Sigma_K(\omega \to 0)$, where the cluster momentum $K$ is $(0,0)$, $(0,\pi)$ and $(\pi,\pi)$. (d) Raw data as a function of hole doping for $\beta=50$. 
(e), (f), (g), (h): Same as in top panels, but for the extrapolated zero-frequency value of the imaginary part of the cluster Cu Green function -Im$G^{d}_K(\omega \to 0)$.}
\label{figS5}
\end{figure*}
\begin{figure*}[!h]
\centering{
\includegraphics[width=0.999\linewidth,clip=]{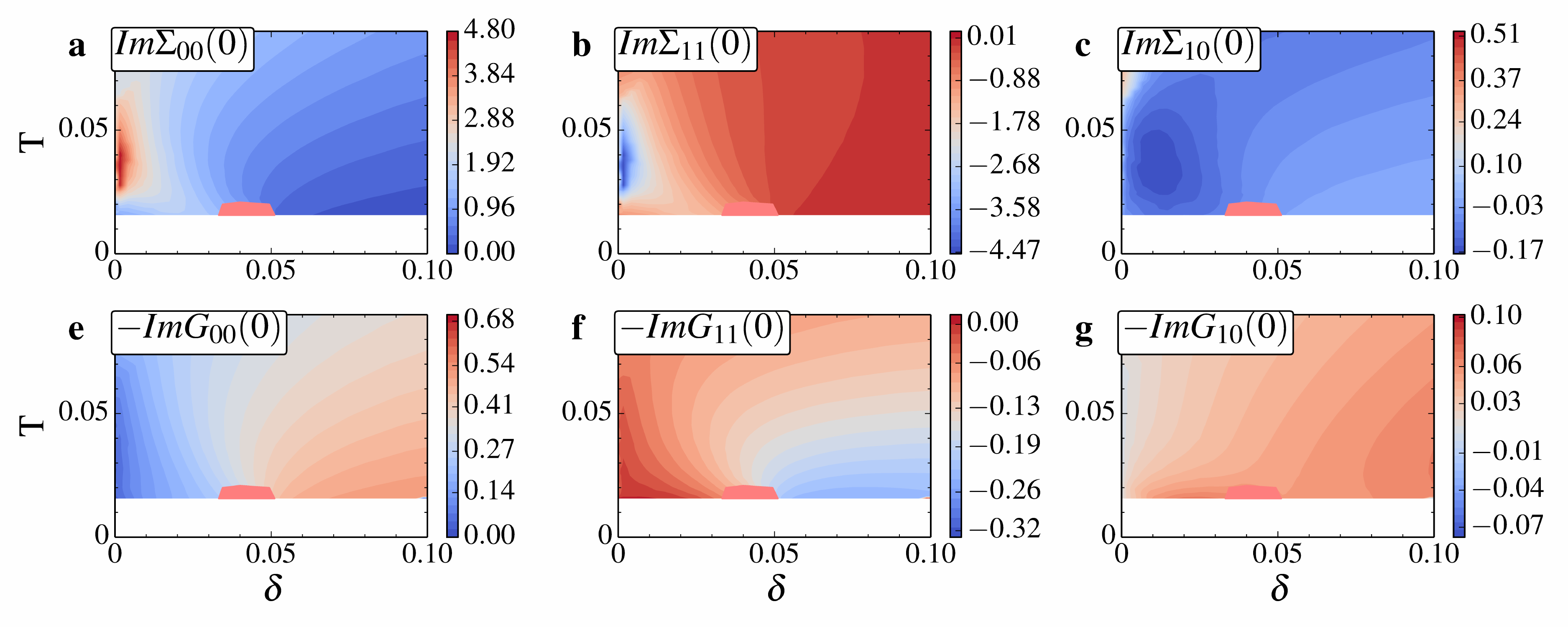}}
\caption{Same graphics as in Fig.~\ref{figS5}, but in real space, where $R=(0,0)$, $(0,1)$ and $(1,1)$. 
}
\label{figS6}
\end{figure*}
\begin{figure*}[!h]
\centering{
\includegraphics[width=0.999\linewidth,clip=]{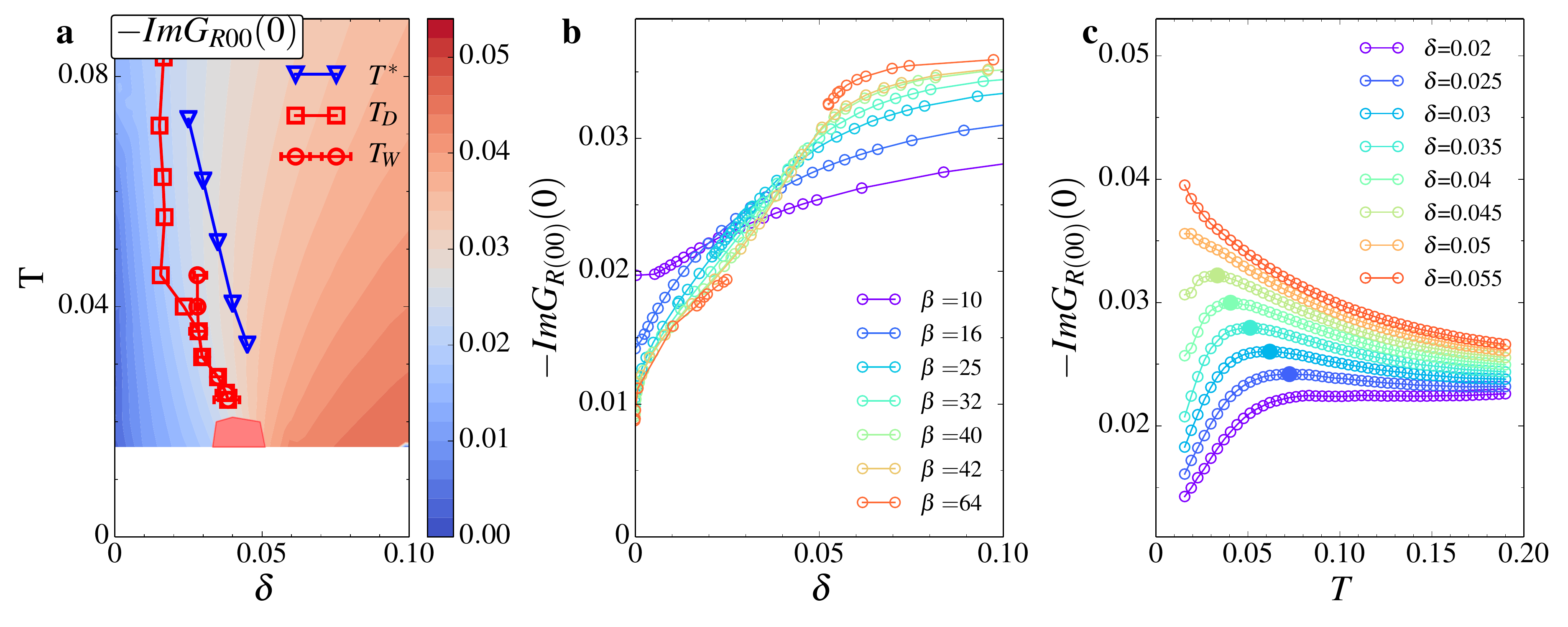}}
\caption{(a) Temperature versus hole doping colormap of the extrapolated zero-frequency value of the imaginary part of the {\it total} cluster Green function -Im$G_{R=(0,0)}(\omega \to 0)$.  Line with red squares shows $T_D$, i.e. the locus of the inflection point of -Im$G_{R=(0,0)}(\omega \to 0)$ as a function of $\mu$. For comparison, $T_W$, i.e.  the locus of charge compressibility maxima, max$_\mu \kappa$, is shown (line with red circles, see also Fig.~3 of main text). As the critical endpoint is approached, these lines become closer. 
(b) Raw data at fixed temperature as a function of hole doping. At the lowest temperature ($\beta=64$), a jump is visible at finite doping.
(c) Raw data at fixed doping as a function of temperature. Filled symbols indicate the onset of $T^*$.  
}
\label{figS7}
\end{figure*}
\begin{figure}[!hb]
\centering{
\includegraphics[width=0.5\linewidth,clip=]{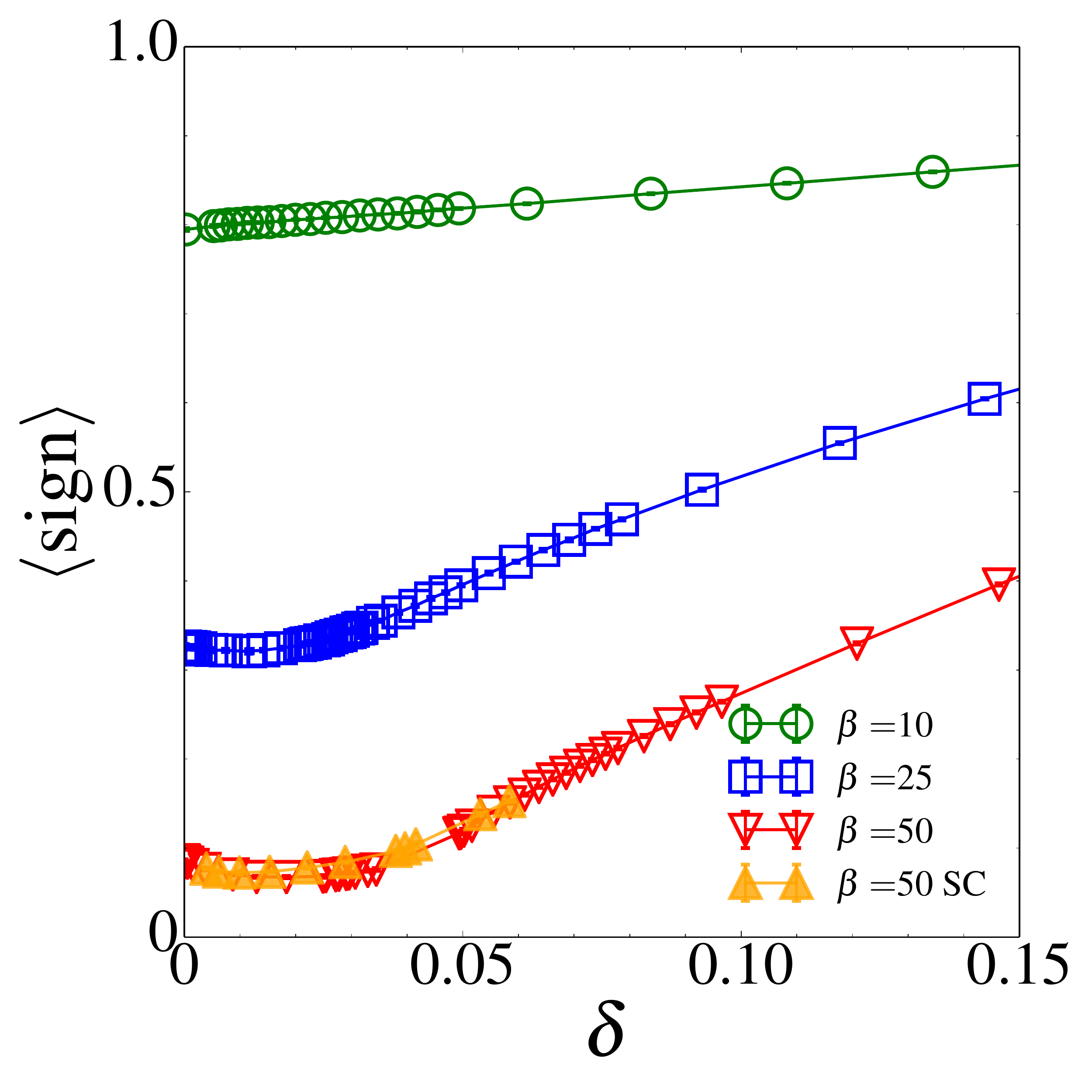}}
\caption{Average sign in CTQMC simulations as a function of hole doping, for $\beta=10, 25$ and $50$. Normal-state data are indicated as open symbols. Superconducting data are shown as full symbols. Other model parameters are $U_d=12$, $|\epsilon_p-\epsilon_d| =9$, $t_{pp}=1$ and $t_{pd}=1.5$.
}
\label{figSX}
\end{figure}
%



\clearpage
\section{Phase characterisation: local density of states}
\label{SectionC}

The four panels of Fig.~4 of main text displayed the partial and total density of states near the Fermi level. It was apparent that over the frequency scale of these figures, the states have comparable copper and oxygen character. The following four figures show the same results but over the whole frequency range. Far below the Fermi level, the states have mostly oxygen character, while they have mostly copper character far above the Fermi level. The copper band that was near $\omega=-11$ at $U_d=0$ becomes incoherent in the charge-transfer insulator regime, spreading over a large frequency range. Note however that the maximum entropy analytic continuation used for these figures~\cite{BergeronMaxEnt:2015_SM} is less reliable far from the Fermi level. 

\vspace{1cm}
\begin{figure}[!hb]
\centering{
\includegraphics[width=0.75\linewidth,clip=]{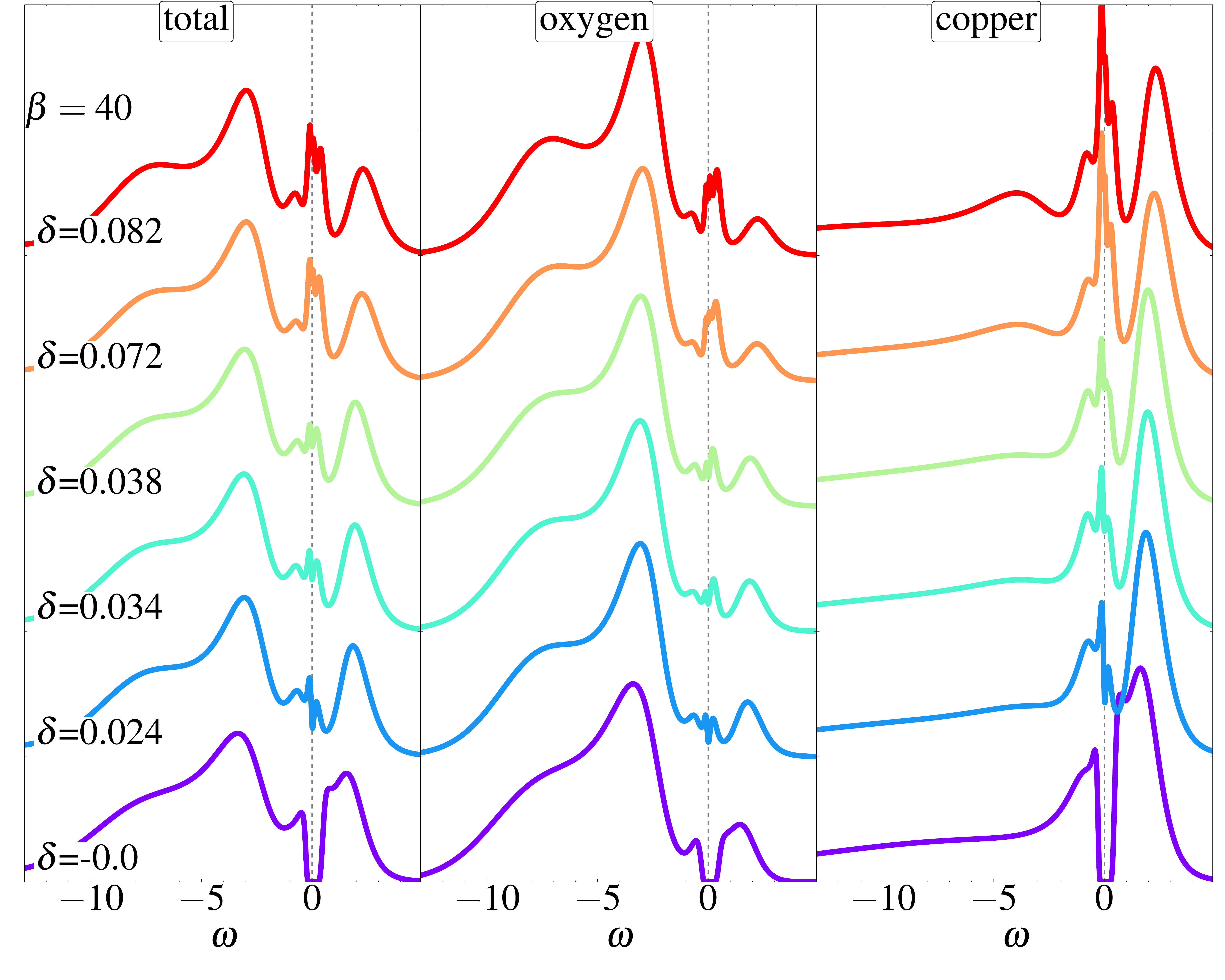}}
\caption{Full frequency spectrum of the DOS shown in Fig.4a of main text.
}
\label{figS8}
\end{figure}
\begin{figure}[!h]
\centering{
\includegraphics[width=0.75\linewidth,clip=]{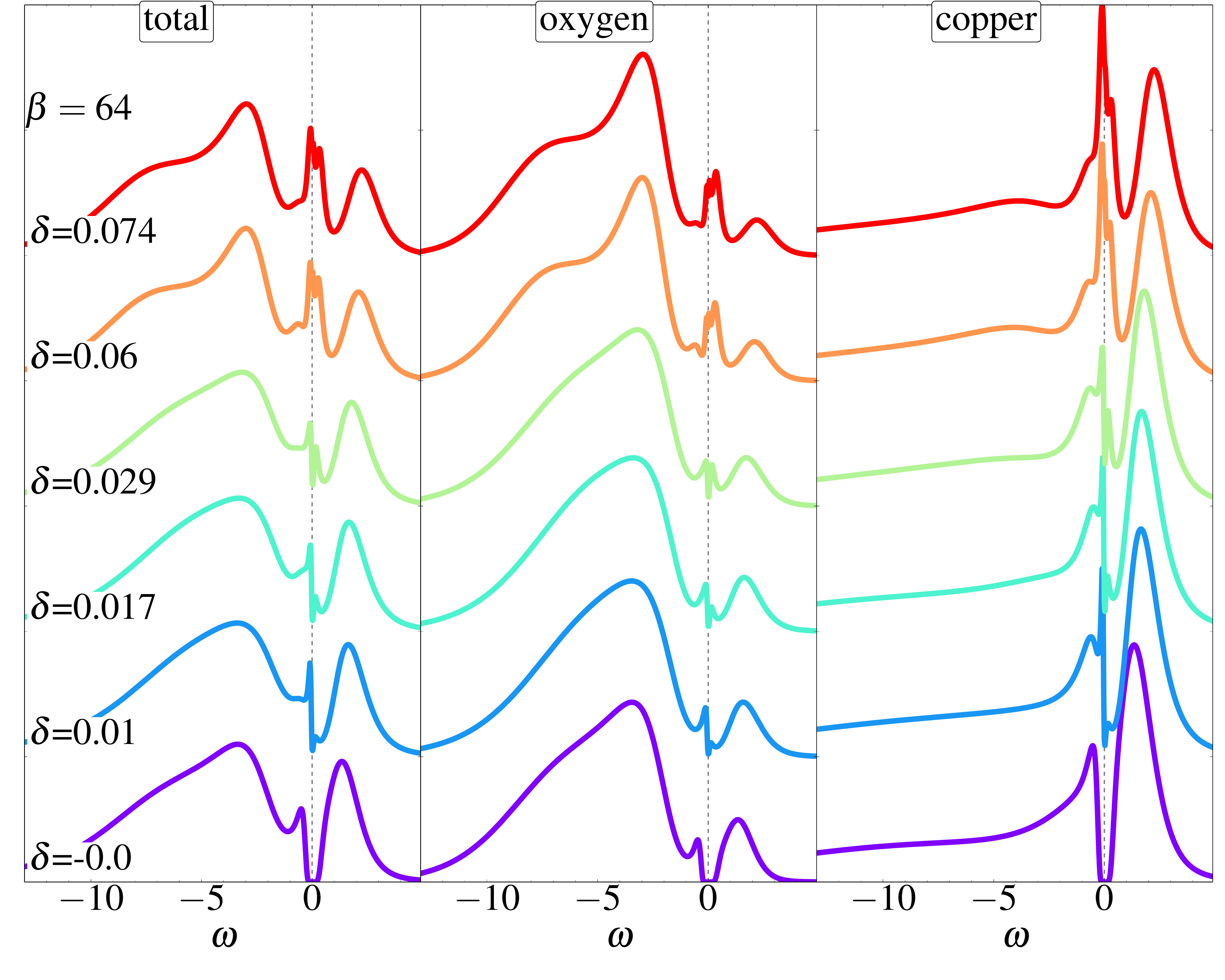}}
\caption{Full frequency spectrum of the DOS shown in Fig.~4b of main text.
}
\label{figS9}
\end{figure}
\begin{figure}[!h]
\centering{
\includegraphics[width=0.75\linewidth,clip=]{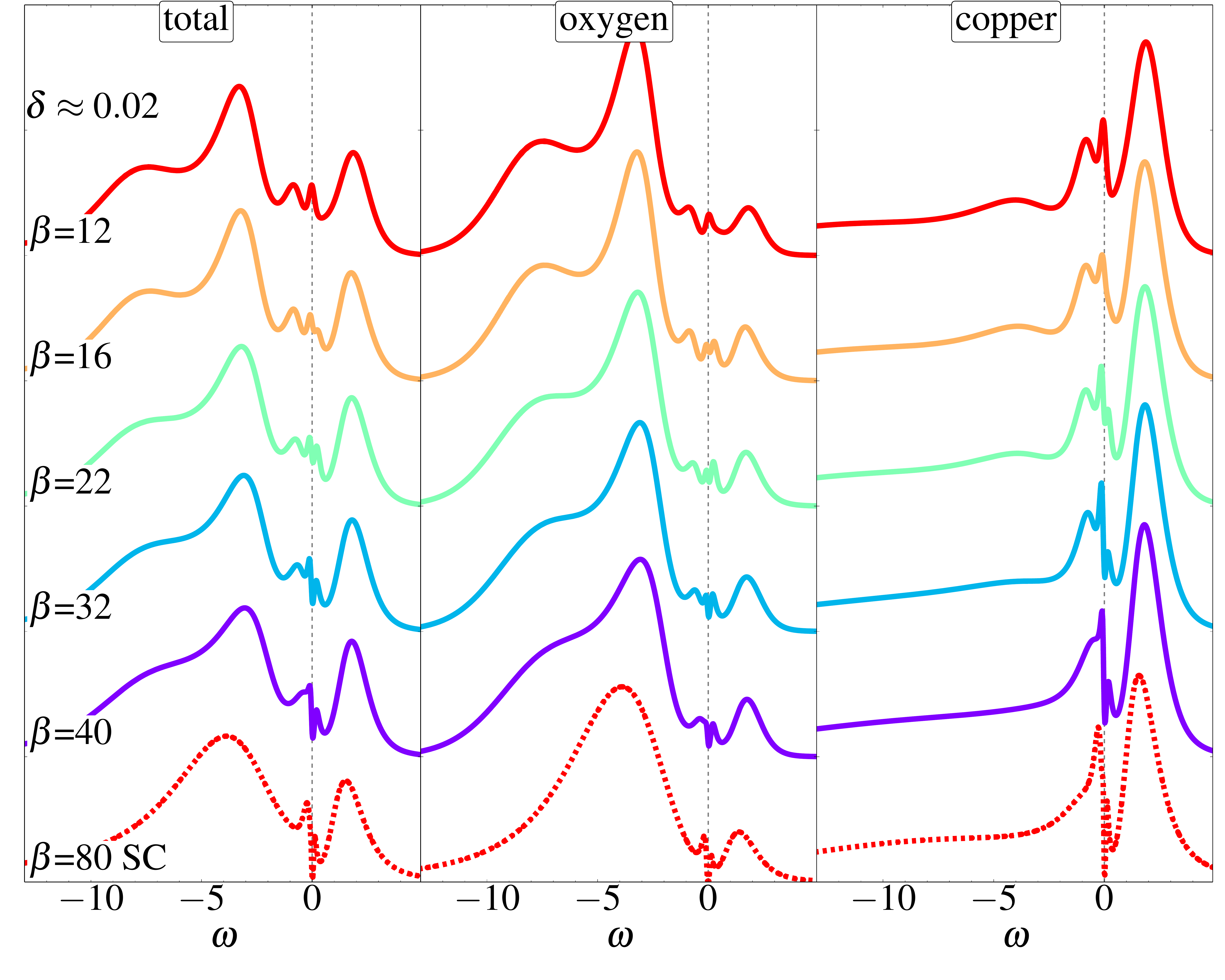}}
\caption{Full frequency spectrum of the DOS shown in Fig.~4c of main text.
}
\label{figS10}
\end{figure}
\begin{figure}
\centering{
\includegraphics[width=0.75\linewidth,clip=]{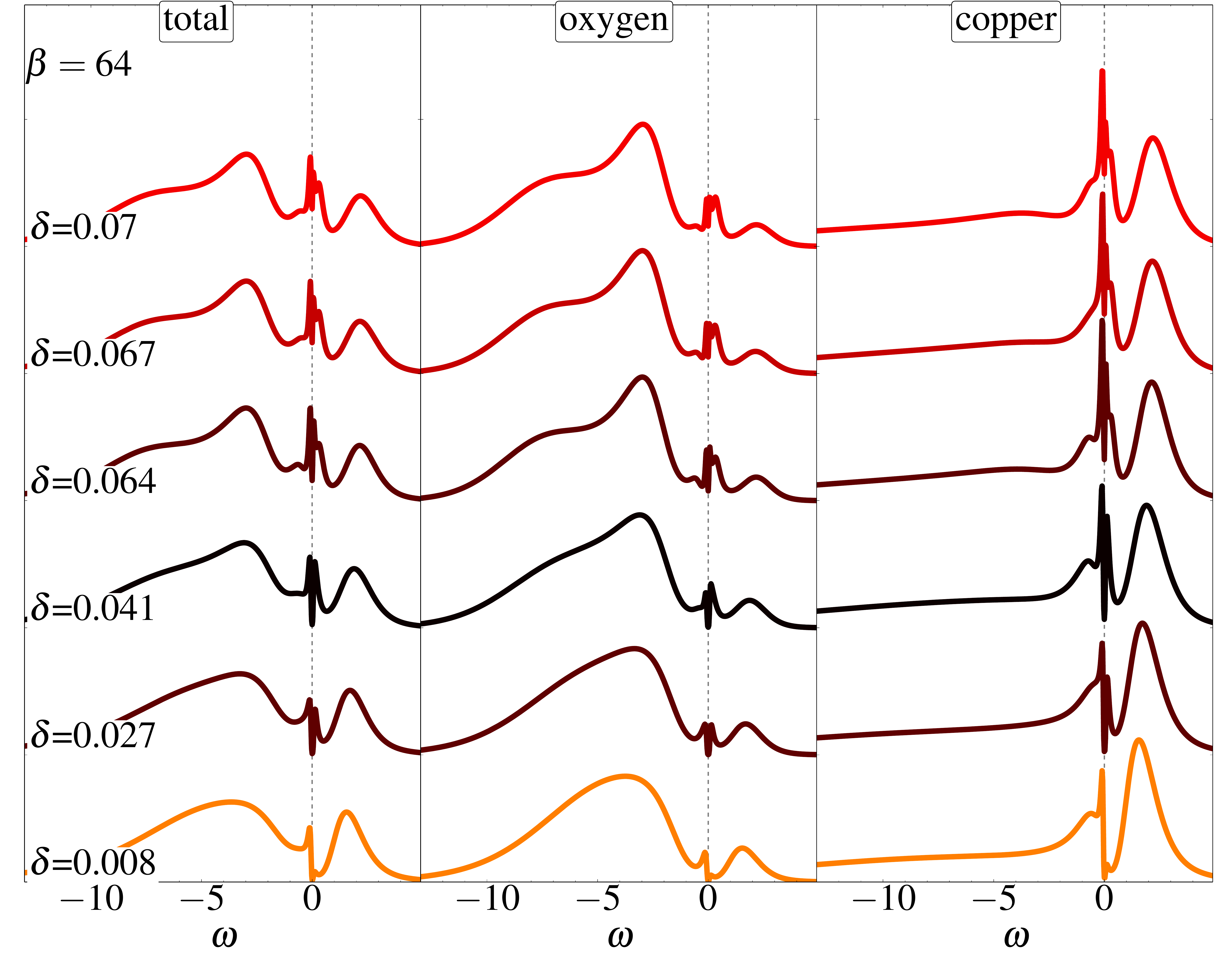}}
\caption{Full frequency spectrum of the DOS shown in Fig.4d of main text.
}
\label{figS11}
\end{figure}
\begin{figure}[!hb]
\centering{
\includegraphics[width=0.55\linewidth,clip=]{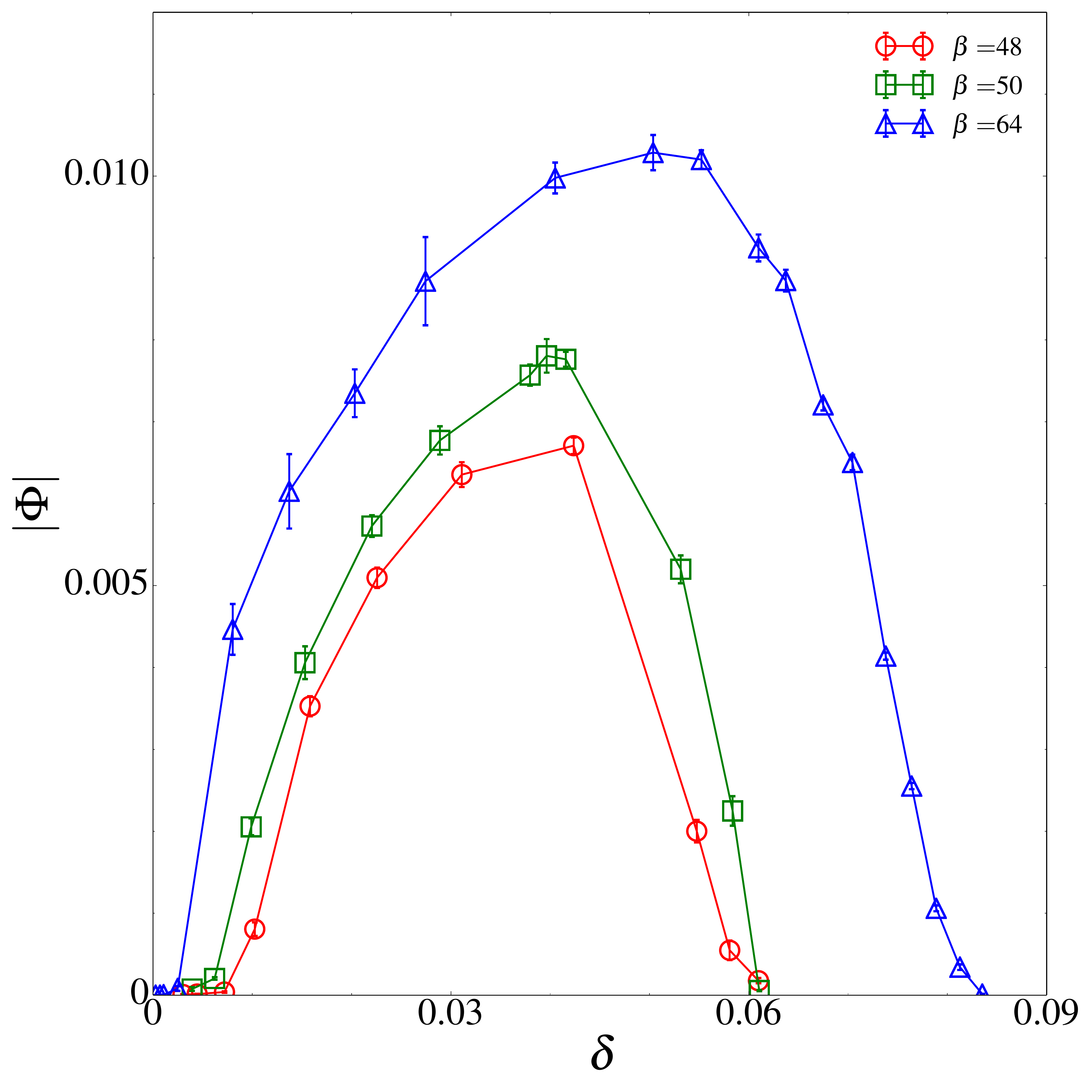}}
\caption{Superconducting order parameter versus $\delta$ for different inverse temperatures $\beta$. The superconducting region in the $T$-$\delta$ phase diagram of Fig.~3 of main text is defined as the region where $\Phi$ is nonzero. 
}
\label{figS12}
\end{figure}
%


\begin{thebibliography}{60}%
\makeatletter
\providecommand \@ifxundefined [1]{%
 \@ifx{#1\undefined}
}%
\providecommand \@ifnum [1]{%
 \ifnum #1\expandafter \@firstoftwo
 \else \expandafter \@secondoftwo
 \fi
}%
\providecommand \@ifx [1]{%
 \ifx #1\expandafter \@firstoftwo
 \else \expandafter \@secondoftwo
 \fi
}%
\providecommand \natexlab [1]{#1}%
\providecommand \enquote  [1]{``#1''}%
\providecommand \bibnamefont  [1]{#1}%
\providecommand \bibfnamefont [1]{#1}%
\providecommand \citenamefont [1]{#1}%
\providecommand \href@noop [0]{\@secondoftwo}%
\providecommand \href [0]{\begingroup \@sanitize@url \@href}%
\providecommand \@href[1]{\@@startlink{#1}\@@href}%
\providecommand \@@href[1]{\endgroup#1\@@endlink}%
\providecommand \@sanitize@url [0]{\catcode `\\12\catcode `\$12\catcode
  `\&12\catcode `\#12\catcode `\^12\catcode `\_12\catcode `\%12\relax}%
\providecommand \@@startlink[1]{}%
\providecommand \@@endlink[0]{}%
\providecommand \url  [0]{\begingroup\@sanitize@url \@url }%
\providecommand \@url [1]{\endgroup\@href {#1}{\urlprefix }}%
\providecommand \urlprefix  [0]{URL }%
\providecommand \Eprint [0]{\href }%
\providecommand \doibase [0]{http://dx.doi.org/}%
\providecommand \selectlanguage [0]{\@gobble}%
\providecommand \bibinfo  [0]{\@secondoftwo}%
\providecommand \bibfield  [0]{\@secondoftwo}%
\providecommand \translation [1]{[#1]}%
\providecommand \BibitemOpen [0]{}%
\providecommand \bibitemStop [0]{}%
\providecommand \bibitemNoStop [0]{.\EOS\space}%
\providecommand \EOS [0]{\spacefactor3000\relax}%
\providecommand \BibitemShut  [1]{\csname bibitem#1\endcsname}%
\let\auto@bib@innerbib\@empty
\bibitem [{\citenamefont {Keimer}\ \emph {et~al.}(2015)\citenamefont {Keimer},
  \citenamefont {Kivelson}, \citenamefont {Norman}, \citenamefont {Uchida},\
  and\ \citenamefont {Zaanen}}]{keimerRev}%
  \BibitemOpen
  \bibfield  {author} {\bibinfo {author} {\bibfnamefont {B.}~\bibnamefont
  {Keimer}}, \bibinfo {author} {\bibfnamefont {S.~A.}\ \bibnamefont
  {Kivelson}}, \bibinfo {author} {\bibfnamefont {M.~R.}\ \bibnamefont
  {Norman}}, \bibinfo {author} {\bibfnamefont {S.}~\bibnamefont {Uchida}}, \
  and\ \bibinfo {author} {\bibfnamefont {J.}~\bibnamefont {Zaanen}},\ }\href
  {\doibase 10.1038/nature14165} {\bibfield  {journal} {\bibinfo  {journal}
  {Nature}\ }\textbf {\bibinfo {volume} {518}},\ \bibinfo {pages} {179}
  (\bibinfo {year} {2015})}\BibitemShut {NoStop}%
\bibitem [{\citenamefont {Anderson}(1987)}]{anderson:1987}%
  \BibitemOpen
  \bibfield  {author} {\bibinfo {author} {\bibfnamefont {P.~W.}\ \bibnamefont
  {Anderson}},\ }\href {\doibase 10.1126/science.235.4793.1196} {\bibfield
  {journal} {\bibinfo  {journal} {Science}\ }\textbf {\bibinfo {volume}
  {235}},\ \bibinfo {pages} {1196} (\bibinfo {year} {1987})}\BibitemShut
  {NoStop}%
\bibitem [{\citenamefont {Maier}\ \emph {et~al.}(2005)\citenamefont {Maier},
  \citenamefont {Jarrell}, \citenamefont {Pruschke},\ and\ \citenamefont
  {Hettler}}]{maier}%
  \BibitemOpen
  \bibfield  {author} {\bibinfo {author} {\bibfnamefont {T.}~\bibnamefont
  {Maier}}, \bibinfo {author} {\bibfnamefont {M.}~\bibnamefont {Jarrell}},
  \bibinfo {author} {\bibfnamefont {T.}~\bibnamefont {Pruschke}}, \ and\
  \bibinfo {author} {\bibfnamefont {M.~H.}\ \bibnamefont {Hettler}},\ }\href
  {\doibase 10.1103/RevModPhys.77.1027} {\bibfield  {journal} {\bibinfo
  {journal} {Rev. Mod. Phys.}\ }\textbf {\bibinfo {volume} {77}},\ \bibinfo
  {pages} {1027} (\bibinfo {year} {2005})}\BibitemShut {NoStop}%
\bibitem [{\citenamefont {Kotliar}\ \emph {et~al.}(2006)\citenamefont
  {Kotliar}, \citenamefont {Savrasov}, \citenamefont {Haule}, \citenamefont
  {Oudovenko}, \citenamefont {Parcollet},\ and\ \citenamefont
  {Marianetti}}]{kotliarRMP}%
  \BibitemOpen
  \bibfield  {author} {\bibinfo {author} {\bibfnamefont {G.}~\bibnamefont
  {Kotliar}}, \bibinfo {author} {\bibfnamefont {S.~Y.}\ \bibnamefont
  {Savrasov}}, \bibinfo {author} {\bibfnamefont {K.}~\bibnamefont {Haule}},
  \bibinfo {author} {\bibfnamefont {V.~S.}\ \bibnamefont {Oudovenko}}, \bibinfo
  {author} {\bibfnamefont {O.}~\bibnamefont {Parcollet}}, \ and\ \bibinfo
  {author} {\bibfnamefont {C.~A.}\ \bibnamefont {Marianetti}},\ }\href
  {\doibase 10.1103/RevModPhys.78.865} {\bibfield  {journal} {\bibinfo
  {journal} {Rev. Mod. Phys.}\ }\textbf {\bibinfo {volume} {78}},\ \bibinfo
  {eid} {865} (\bibinfo {year} {2006})}\BibitemShut {NoStop}%
\bibitem [{\citenamefont {Tremblay}\ \emph {et~al.}(2006)\citenamefont
  {Tremblay}, \citenamefont {Kyung},\ and\ \citenamefont
  {S\'{e}n\'{e}chal}}]{tremblayR}%
  \BibitemOpen
  \bibfield  {author} {\bibinfo {author} {\bibfnamefont {A.-M.~S.}\
  \bibnamefont {Tremblay}}, \bibinfo {author} {\bibfnamefont {B.}~\bibnamefont
  {Kyung}}, \ and\ \bibinfo {author} {\bibfnamefont {D.}~\bibnamefont
  {S\'{e}n\'{e}chal}},\ }\href {\doibase 10.1063/1.2199446} {\bibfield
  {journal} {\bibinfo  {journal} {Low Temp. Phys.}\ }\textbf {\bibinfo {volume}
  {32}},\ \bibinfo {pages} {424} (\bibinfo {year} {2006})}\BibitemShut
  {NoStop}%
\bibitem [{\citenamefont {Georges}\ \emph {et~al.}(1996)\citenamefont
  {Georges}, \citenamefont {Kotliar}, \citenamefont {Krauth},\ and\
  \citenamefont {Rozenberg}}]{rmp}%
  \BibitemOpen
  \bibfield  {author} {\bibinfo {author} {\bibfnamefont {A.}~\bibnamefont
  {Georges}}, \bibinfo {author} {\bibfnamefont {G.}~\bibnamefont {Kotliar}},
  \bibinfo {author} {\bibfnamefont {W.}~\bibnamefont {Krauth}}, \ and\ \bibinfo
  {author} {\bibfnamefont {M.~J.}\ \bibnamefont {Rozenberg}},\ }\href {\doibase
  10.1103/RevModPhys.68.13} {\bibfield  {journal} {\bibinfo  {journal} {Rev.
  Mod. Phys.}\ }\textbf {\bibinfo {volume} {68}},\ \bibinfo {pages} {13}
  (\bibinfo {year} {1996})}\BibitemShut {NoStop}%
\bibitem [{\citenamefont {Gull}\ and\ \citenamefont
  {Millis}(2015)}]{GullNewsViews:2015}%
  \BibitemOpen
  \bibfield  {author} {\bibinfo {author} {\bibfnamefont {E.}~\bibnamefont
  {Gull}}\ and\ \bibinfo {author} {\bibfnamefont {A.~J.}\ \bibnamefont
  {Millis}},\ }\href {\doibase 10.1038/nphys3501} {\bibfield  {journal}
  {\bibinfo  {journal} {Nat Phys}\ }\textbf {\bibinfo {volume} {11}},\ \bibinfo
  {pages} {808} (\bibinfo {year} {2015})}\BibitemShut {NoStop}%
\bibitem [{\citenamefont {Tremblay}(2013)}]{AMJulich}%
  \BibitemOpen
  \bibfield  {author} {\bibinfo {author} {\bibfnamefont {A.-M.~S.}\
  \bibnamefont {Tremblay}},\ }in\ \href
  {http://juser.fz-juelich.de/record/137827/files/FZJ-2013-04137.pdf?version=1}
  {\emph {\bibinfo {booktitle} {Emergent Phenomena in Correlated Matter
  Modeling and Simulation}}},\ Vol.~\bibinfo {volume} {3},\ \bibinfo {editor}
  {edited by\ \bibinfo {editor} {\bibfnamefont {E.}~\bibnamefont {Pavarini}},
  \bibinfo {editor} {\bibfnamefont {E.}~\bibnamefont {Koch}}, \ and\ \bibinfo
  {editor} {\bibfnamefont {U.}~\bibnamefont {Schollw\"ock}}}\ (\bibinfo
  {publisher} {Verlag des Forschungszentrum},\ \bibinfo {year} {2013})\
  Chap.~\bibinfo {chapter} {10}\BibitemShut {NoStop}%
\bibitem [{\citenamefont {Emery}(1987)}]{Emery_1987}%
  \BibitemOpen
  \bibfield  {author} {\bibinfo {author} {\bibfnamefont {V.~J.}\ \bibnamefont
  {Emery}},\ }\href {\doibase 10.1103/PhysRevLett.58.2794} {\bibfield
  {journal} {\bibinfo  {journal} {Phys. Rev. Lett.}\ }\textbf {\bibinfo
  {volume} {58}},\ \bibinfo {pages} {2794} (\bibinfo {year}
  {1987})}\BibitemShut {NoStop}%
\bibitem [{\citenamefont {Varma}\ \emph {et~al.}(1987)\citenamefont {Varma},
  \citenamefont {Schmitt-Rink},\ and\ \citenamefont {Abrahams}}]{Varma_1987}%
  \BibitemOpen
  \bibfield  {author} {\bibinfo {author} {\bibfnamefont {C.}~\bibnamefont
  {Varma}}, \bibinfo {author} {\bibfnamefont {S.}~\bibnamefont {Schmitt-Rink}},
  \ and\ \bibinfo {author} {\bibfnamefont {E.}~\bibnamefont {Abrahams}},\
  }\href {http://dx.doi.org/10.1016/0038-1098(87)90407-8} {\bibfield  {journal}
  {\bibinfo  {journal} {Solid State Communications}\ }\textbf {\bibinfo
  {volume} {62}},\ \bibinfo {pages} {681 } (\bibinfo {year}
  {1987})}\BibitemShut {NoStop}%
\bibitem [{\citenamefont {Gauquelin}\ \emph {et~al.}(2014)\citenamefont
  {Gauquelin}, \citenamefont {Hawthorn}, \citenamefont {Sawatzky},
  \citenamefont {Liang}, \citenamefont {Bonn}, \citenamefont {Hardy},\ and\
  \citenamefont {Botton}}]{Gauquelin2014}%
  \BibitemOpen
  \bibfield  {author} {\bibinfo {author} {\bibfnamefont {N.}~\bibnamefont
  {Gauquelin}}, \bibinfo {author} {\bibfnamefont {D.}~\bibnamefont {Hawthorn}},
  \bibinfo {author} {\bibfnamefont {G.}~\bibnamefont {Sawatzky}}, \bibinfo
  {author} {\bibfnamefont {R.}~\bibnamefont {Liang}}, \bibinfo {author}
  {\bibfnamefont {D.}~\bibnamefont {Bonn}}, \bibinfo {author} {\bibfnamefont
  {W.}~\bibnamefont {Hardy}}, \ and\ \bibinfo {author} {\bibfnamefont
  {G.}~\bibnamefont {Botton}},\ }\href {\doibase 10.1038/ncomms5275} {\bibfield
   {journal} {\bibinfo  {journal} {Nature Communications}\ }\textbf {\bibinfo
  {volume} {5}} (\bibinfo {year} {2014}),\ 10.1038/ncomms5275}\BibitemShut
  {NoStop}%
\bibitem [{\citenamefont {Zaanen}\ \emph {et~al.}(1985)\citenamefont {Zaanen},
  \citenamefont {Sawatzky},\ and\ \citenamefont {Allen}}]{zsa}%
  \BibitemOpen
  \bibfield  {author} {\bibinfo {author} {\bibfnamefont {J.}~\bibnamefont
  {Zaanen}}, \bibinfo {author} {\bibfnamefont {G.~A.}\ \bibnamefont
  {Sawatzky}}, \ and\ \bibinfo {author} {\bibfnamefont {J.~W.}\ \bibnamefont
  {Allen}},\ }\href {\doibase 10.1103/PhysRevLett.55.418} {\bibfield  {journal}
  {\bibinfo  {journal} {Phys. Rev. Lett.}\ }\textbf {\bibinfo {volume} {55}},\
  \bibinfo {pages} {418} (\bibinfo {year} {1985})}\BibitemShut {NoStop}%
\bibitem [{\citenamefont {Georges}\ \emph {et~al.}(1993)\citenamefont
  {Georges}, \citenamefont {Kotliar},\ and\ \citenamefont
  {Krauth}}]{AntoineCuO2}%
  \BibitemOpen
  \bibfield  {author} {\bibinfo {author} {\bibfnamefont {A.}~\bibnamefont
  {Georges}}, \bibinfo {author} {\bibfnamefont {G.}~\bibnamefont {Kotliar}}, \
  and\ \bibinfo {author} {\bibfnamefont {W.}~\bibnamefont {Krauth}},\ }\href
  {\doibase 10.1007/BF01308748} {\bibfield  {journal} {\bibinfo  {journal}
  {Zeitschrift f{\"u}r Physik B Condensed Matter}\ }\textbf {\bibinfo {volume}
  {92}},\ \bibinfo {pages} {313} (\bibinfo {year} {1993})}\BibitemShut
  {NoStop}%
\bibitem [{\citenamefont {Lombardo}\ \emph {et~al.}(1996)\citenamefont
  {Lombardo}, \citenamefont {Avignon}, \citenamefont {Schmalian},\ and\
  \citenamefont {Bennemann}}]{Lombardo1996}%
  \BibitemOpen
  \bibfield  {author} {\bibinfo {author} {\bibfnamefont {P.}~\bibnamefont
  {Lombardo}}, \bibinfo {author} {\bibfnamefont {M.}~\bibnamefont {Avignon}},
  \bibinfo {author} {\bibfnamefont {J.}~\bibnamefont {Schmalian}}, \ and\
  \bibinfo {author} {\bibfnamefont {K.-H.}\ \bibnamefont {Bennemann}},\ }\href
  {\doibase 10.1103/PhysRevB.54.5317} {\bibfield  {journal} {\bibinfo
  {journal} {Phys. Rev. B}\ }\textbf {\bibinfo {volume} {54}},\ \bibinfo
  {pages} {5317} (\bibinfo {year} {1996})}\BibitemShut {NoStop}%
\bibitem [{\citenamefont {Z{\"o}lfl}\ \emph {et~al.}(2000)\citenamefont
  {Z{\"o}lfl}, \citenamefont {Maier}, \citenamefont {Pruschke},\ and\
  \citenamefont {Keller}}]{Zolf2000}%
  \BibitemOpen
  \bibfield  {author} {\bibinfo {author} {\bibfnamefont {M.}~\bibnamefont
  {Z{\"o}lfl}}, \bibinfo {author} {\bibfnamefont {T.}~\bibnamefont {Maier}},
  \bibinfo {author} {\bibfnamefont {T.}~\bibnamefont {Pruschke}}, \ and\
  \bibinfo {author} {\bibfnamefont {J.}~\bibnamefont {Keller}},\ }\href
  {\doibase 10.1007/s100510050009} {\bibfield  {journal} {\bibinfo  {journal}
  {The European Physical Journal B - Condensed Matter and Complex Systems}\
  }\textbf {\bibinfo {volume} {13}},\ \bibinfo {pages} {47} (\bibinfo {year}
  {2000})}\BibitemShut {NoStop}%
\bibitem [{\citenamefont {Sordi}\ \emph {et~al.}(2007)\citenamefont {Sordi},
  \citenamefont {Amaricci},\ and\ \citenamefont {Rozenberg}}]{sar}%
  \BibitemOpen
  \bibfield  {author} {\bibinfo {author} {\bibfnamefont {G.}~\bibnamefont
  {Sordi}}, \bibinfo {author} {\bibfnamefont {A.}~\bibnamefont {Amaricci}}, \
  and\ \bibinfo {author} {\bibfnamefont {M.~J.}\ \bibnamefont {Rozenberg}},\
  }\href {\doibase 10.1103/PhysRevLett.99.196403} {\bibfield  {journal}
  {\bibinfo  {journal} {Phys. Rev. Lett.}\ }\textbf {\bibinfo {volume} {99}},\
  \bibinfo {pages} {196403} (\bibinfo {year} {2007})}\BibitemShut {NoStop}%
\bibitem [{\citenamefont {Weber}\ \emph {et~al.}(2008)\citenamefont {Weber},
  \citenamefont {Haule},\ and\ \citenamefont {Kotliar}}]{Cedric2008}%
  \BibitemOpen
  \bibfield  {author} {\bibinfo {author} {\bibfnamefont {C.}~\bibnamefont
  {Weber}}, \bibinfo {author} {\bibfnamefont {K.}~\bibnamefont {Haule}}, \ and\
  \bibinfo {author} {\bibfnamefont {G.}~\bibnamefont {Kotliar}},\ }\href
  {\doibase 10.1103/PhysRevB.78.134519} {\bibfield  {journal} {\bibinfo
  {journal} {Phys. Rev. B}\ }\textbf {\bibinfo {volume} {78}},\ \bibinfo
  {pages} {134519} (\bibinfo {year} {2008})}\BibitemShut {NoStop}%
\bibitem [{\citenamefont {Weber}\ \emph {et~al.}(2010)\citenamefont {Weber},
  \citenamefont {Haule},\ and\ \citenamefont {Kotliar}}]{Weber:2010}%
  \BibitemOpen
  \bibfield  {author} {\bibinfo {author} {\bibfnamefont {C.}~\bibnamefont
  {Weber}}, \bibinfo {author} {\bibfnamefont {K.}~\bibnamefont {Haule}}, \ and\
  \bibinfo {author} {\bibfnamefont {G.}~\bibnamefont {Kotliar}},\ }\href
  {\doibase 10.1038/nphys1706} {\bibfield  {journal} {\bibinfo  {journal}
  {Nature Physics}\ }\textbf {\bibinfo {volume} {6}},\ \bibinfo {pages} {574}
  (\bibinfo {year} {2010})}\BibitemShut {NoStop}%
\bibitem [{\citenamefont {de' Medici}\ \emph {et~al.}(2009)\citenamefont {de'
  Medici}, \citenamefont {Wang}, \citenamefont {Capone},\ and\ \citenamefont
  {Millis}}]{DeMedici2009}%
  \BibitemOpen
  \bibfield  {author} {\bibinfo {author} {\bibfnamefont {L.}~\bibnamefont {de'
  Medici}}, \bibinfo {author} {\bibfnamefont {X.}~\bibnamefont {Wang}},
  \bibinfo {author} {\bibfnamefont {M.}~\bibnamefont {Capone}}, \ and\ \bibinfo
  {author} {\bibfnamefont {A.~J.}\ \bibnamefont {Millis}},\ }\href {\doibase
  10.1103/PhysRevB.80.054501} {\bibfield  {journal} {\bibinfo  {journal} {Phys.
  Rev. B}\ }\textbf {\bibinfo {volume} {80}},\ \bibinfo {pages} {054501}
  (\bibinfo {year} {2009})}\BibitemShut {NoStop}%
\bibitem [{\citenamefont {Wang}\ \emph
  {et~al.}(2011{\natexlab{a}})\citenamefont {Wang}, \citenamefont {de'
  Medici},\ and\ \citenamefont {Millis}}]{Wang:2011}%
  \BibitemOpen
  \bibfield  {author} {\bibinfo {author} {\bibfnamefont {X.}~\bibnamefont
  {Wang}}, \bibinfo {author} {\bibfnamefont {L.}~\bibnamefont {de' Medici}}, \
  and\ \bibinfo {author} {\bibfnamefont {A.~J.}\ \bibnamefont {Millis}},\
  }\href {\doibase 10.1103/PhysRevB.83.094501} {\bibfield  {journal} {\bibinfo
  {journal} {Phys. Rev. B}\ }\textbf {\bibinfo {volume} {83}},\ \bibinfo
  {pages} {094501} (\bibinfo {year} {2011}{\natexlab{a}})}\BibitemShut
  {NoStop}%
\bibitem [{\citenamefont {Wang}\ \emph
  {et~al.}(2011{\natexlab{b}})\citenamefont {Wang}, \citenamefont {Dang},\ and\
  \citenamefont {Millis}}]{Wang:2011b}%
  \BibitemOpen
  \bibfield  {author} {\bibinfo {author} {\bibfnamefont {X.}~\bibnamefont
  {Wang}}, \bibinfo {author} {\bibfnamefont {H.~T.}\ \bibnamefont {Dang}}, \
  and\ \bibinfo {author} {\bibfnamefont {A.~J.}\ \bibnamefont {Millis}},\
  }\href {\doibase 10.1103/PhysRevB.84.014530} {\bibfield  {journal} {\bibinfo
  {journal} {Phys. Rev. B}\ }\textbf {\bibinfo {volume} {84}},\ \bibinfo
  {pages} {014530} (\bibinfo {year} {2011}{\natexlab{b}})}\BibitemShut
  {NoStop}%
\bibitem [{\citenamefont {Macridin}\ \emph {et~al.}(2005)\citenamefont
  {Macridin}, \citenamefont {Jarrell}, \citenamefont {Maier},\ and\
  \citenamefont {Sawatzky}}]{Macridin2005}%
  \BibitemOpen
  \bibfield  {author} {\bibinfo {author} {\bibfnamefont {A.}~\bibnamefont
  {Macridin}}, \bibinfo {author} {\bibfnamefont {M.}~\bibnamefont {Jarrell}},
  \bibinfo {author} {\bibfnamefont {T.}~\bibnamefont {Maier}}, \ and\ \bibinfo
  {author} {\bibfnamefont {G.~A.}\ \bibnamefont {Sawatzky}},\ }\href {\doibase
  10.1103/PhysRevB.71.134527} {\bibfield  {journal} {\bibinfo  {journal} {Phys.
  Rev. B}\ }\textbf {\bibinfo {volume} {71}},\ \bibinfo {pages} {134527}
  (\bibinfo {year} {2005})}\BibitemShut {NoStop}%
\bibitem [{\citenamefont {Arrigoni}\ \emph {et~al.}(2009)\citenamefont
  {Arrigoni}, \citenamefont {Aichhorn}, \citenamefont {Daghofer},\ and\
  \citenamefont {Hanke}}]{ArrigoniCuO2}%
  \BibitemOpen
  \bibfield  {author} {\bibinfo {author} {\bibfnamefont {E.}~\bibnamefont
  {Arrigoni}}, \bibinfo {author} {\bibfnamefont {M.}~\bibnamefont {Aichhorn}},
  \bibinfo {author} {\bibfnamefont {M.}~\bibnamefont {Daghofer}}, \ and\
  \bibinfo {author} {\bibfnamefont {W.}~\bibnamefont {Hanke}},\ }\href
  {http://stacks.iop.org/1367-2630/11/i=5/a=055066} {\bibfield  {journal}
  {\bibinfo  {journal} {New Journal of Physics}\ }\textbf {\bibinfo {volume}
  {11}},\ \bibinfo {pages} {055066} (\bibinfo {year} {2009})}\BibitemShut
  {NoStop}%
\bibitem [{\citenamefont {Weber}\ \emph {et~al.}(2012)\citenamefont {Weber},
  \citenamefont {Yee}, \citenamefont {Haule},\ and\ \citenamefont
  {Kotliar}}]{Weber2011}%
  \BibitemOpen
  \bibfield  {author} {\bibinfo {author} {\bibfnamefont {C.}~\bibnamefont
  {Weber}}, \bibinfo {author} {\bibfnamefont {C.}~\bibnamefont {Yee}}, \bibinfo
  {author} {\bibfnamefont {K.}~\bibnamefont {Haule}}, \ and\ \bibinfo {author}
  {\bibfnamefont {G.}~\bibnamefont {Kotliar}},\ }\href
  {http://stacks.iop.org/0295-5075/100/i=3/a=37001} {\bibfield  {journal}
  {\bibinfo  {journal} {EPL (Europhysics Letters)}\ }\textbf {\bibinfo {volume}
  {100}},\ \bibinfo {pages} {37001} (\bibinfo {year} {2012})}\BibitemShut
  {NoStop}%
\bibitem [{\citenamefont {Go}\ and\ \citenamefont {Millis}(2015)}]{go}%
  \BibitemOpen
  \bibfield  {author} {\bibinfo {author} {\bibfnamefont {A.}~\bibnamefont
  {Go}}\ and\ \bibinfo {author} {\bibfnamefont {A.~J.}\ \bibnamefont
  {Millis}},\ }\href {\doibase 10.1103/PhysRevLett.114.016402} {\bibfield
  {journal} {\bibinfo  {journal} {Phys. Rev. Lett.}\ }\textbf {\bibinfo
  {volume} {114}},\ \bibinfo {pages} {016402} (\bibinfo {year}
  {2015})}\BibitemShut {NoStop}%
\bibitem [{\citenamefont {Andersen}\ \emph {et~al.}(1995)\citenamefont
  {Andersen}, \citenamefont {Liechtenstein}, \citenamefont {Jepsen},\ and\
  \citenamefont {Paulsen}}]{AndersenLDA}%
  \BibitemOpen
  \bibfield  {author} {\bibinfo {author} {\bibfnamefont {O.}~\bibnamefont
  {Andersen}}, \bibinfo {author} {\bibfnamefont {A.}~\bibnamefont
  {Liechtenstein}}, \bibinfo {author} {\bibfnamefont {O.}~\bibnamefont
  {Jepsen}}, \ and\ \bibinfo {author} {\bibfnamefont {F.}~\bibnamefont
  {Paulsen}},\ }\href {\doibase http://dx.doi.org/10.1016/0022-3697(95)00269-3}
  {\bibfield  {journal} {\bibinfo  {journal} {J. Phys, Chem. Solids}\ }\textbf
  {\bibinfo {volume} {56}},\ \bibinfo {pages} {1573} (\bibinfo {year}
  {1995})}\BibitemShut {NoStop}%
\bibitem [{\citenamefont {Gull}\ \emph {et~al.}(2011)\citenamefont {Gull},
  \citenamefont {Millis}, \citenamefont {Lichtenstein}, \citenamefont
  {Rubtsov}, \citenamefont {Troyer},\ and\ \citenamefont {Werner}}]{millisRMP}%
  \BibitemOpen
  \bibfield  {author} {\bibinfo {author} {\bibfnamefont {E.}~\bibnamefont
  {Gull}}, \bibinfo {author} {\bibfnamefont {A.~J.}\ \bibnamefont {Millis}},
  \bibinfo {author} {\bibfnamefont {A.~I.}\ \bibnamefont {Lichtenstein}},
  \bibinfo {author} {\bibfnamefont {A.~N.}\ \bibnamefont {Rubtsov}}, \bibinfo
  {author} {\bibfnamefont {M.}~\bibnamefont {Troyer}}, \ and\ \bibinfo {author}
  {\bibfnamefont {P.}~\bibnamefont {Werner}},\ }\href {\doibase
  10.1103/RevModPhys.83.349} {\bibfield  {journal} {\bibinfo  {journal} {Rev.
  Mod. Phys.}\ }\textbf {\bibinfo {volume} {83}},\ \bibinfo {pages} {349}
  (\bibinfo {year} {2011})}\BibitemShut {NoStop}%
\bibitem [{\citenamefont {Xu}\ \emph {et~al.}(2005)\citenamefont {Xu},
  \citenamefont {Kumar}, \citenamefont {Buldyrev}, \citenamefont {Chen},
  \citenamefont {Poole}, \citenamefont {Sciortino},\ and\ \citenamefont
  {Stanley}}]{water1}%
  \BibitemOpen
  \bibfield  {author} {\bibinfo {author} {\bibfnamefont {L.}~\bibnamefont
  {Xu}}, \bibinfo {author} {\bibfnamefont {P.}~\bibnamefont {Kumar}}, \bibinfo
  {author} {\bibfnamefont {S.~V.}\ \bibnamefont {Buldyrev}}, \bibinfo {author}
  {\bibfnamefont {S.-H.}\ \bibnamefont {Chen}}, \bibinfo {author}
  {\bibfnamefont {P.~H.}\ \bibnamefont {Poole}}, \bibinfo {author}
  {\bibfnamefont {F.}~\bibnamefont {Sciortino}}, \ and\ \bibinfo {author}
  {\bibfnamefont {H.~E.}\ \bibnamefont {Stanley}},\ }\href {\doibase
  10.1073/pnas.0507870102} {\bibfield  {journal} {\bibinfo  {journal} {Proc.
  Natl. Acad. Sci. USA}\ }\textbf {\bibinfo {volume} {102}},\ \bibinfo {pages}
  {16558} (\bibinfo {year} {2005})}\BibitemShut {NoStop}%
\bibitem [{\citenamefont {Sordi}\ \emph
  {et~al.}(2012{\natexlab{a}})\citenamefont {Sordi}, \citenamefont {S\'emon},
  \citenamefont {Haule},\ and\ \citenamefont {Tremblay}}]{ssht}%
  \BibitemOpen
  \bibfield  {author} {\bibinfo {author} {\bibfnamefont {G.}~\bibnamefont
  {Sordi}}, \bibinfo {author} {\bibfnamefont {P.}~\bibnamefont {S\'emon}},
  \bibinfo {author} {\bibfnamefont {K.}~\bibnamefont {Haule}}, \ and\ \bibinfo
  {author} {\bibfnamefont {A.-M.~S.}\ \bibnamefont {Tremblay}},\ }\href
  {\doibase doi:10.1038/srep00547} {\bibfield  {journal} {\bibinfo  {journal}
  {Sci. Rep.}\ }\textbf {\bibinfo {volume} {2}},\ \bibinfo {pages} {547}
  (\bibinfo {year} {2012}{\natexlab{a}})}\BibitemShut {NoStop}%
\bibitem [{\citenamefont {Alloul}(2014)}]{Alloul2013}%
  \BibitemOpen
  \bibfield  {author} {\bibinfo {author} {\bibfnamefont {H.}~\bibnamefont
  {Alloul}},\ }\href {\doibase 10.1016/j.crhy.2014.02.007} {\bibfield
  {journal} {\bibinfo  {journal} {Comptes Rendus Physique}\ }\textbf {\bibinfo
  {volume} {15}},\ \bibinfo {pages} {519 } (\bibinfo {year}
  {2014})}\BibitemShut {NoStop}%
\bibitem [{\citenamefont {Molegraaf}\ \emph {et~al.}(2002)\citenamefont
  {Molegraaf}, \citenamefont {Presura}, \citenamefont {van~der Marel},
  \citenamefont {Kes},\ and\ \citenamefont {Li}}]{Molegraaf2002}%
  \BibitemOpen
  \bibfield  {author} {\bibinfo {author} {\bibfnamefont {H.~J.~A.}\
  \bibnamefont {Molegraaf}}, \bibinfo {author} {\bibfnamefont {C.}~\bibnamefont
  {Presura}}, \bibinfo {author} {\bibfnamefont {D.}~\bibnamefont {van~der
  Marel}}, \bibinfo {author} {\bibfnamefont {P.~H.}\ \bibnamefont {Kes}}, \
  and\ \bibinfo {author} {\bibfnamefont {M.}~\bibnamefont {Li}},\ }\href
  {\doibase 10.1126/science.1069947} {\bibfield  {journal} {\bibinfo  {journal}
  {Science}\ }\textbf {\bibinfo {volume} {295}},\ \bibinfo {pages} {2239}
  (\bibinfo {year} {2002})}\BibitemShut {NoStop}%
\bibitem [{\citenamefont {{Giannetti}}\ \emph {et~al.}(2011)\citenamefont
  {{Giannetti}}, \citenamefont {{Cilento}}, \citenamefont {{Conte}},
  \citenamefont {{Coslovich}}, \citenamefont {{Ferrini}}, \citenamefont
  {{Molegraaf}}, \citenamefont {{Raichle}}, \citenamefont {{Liang}},
  \citenamefont {{Eisaki}}, \citenamefont {{Greven}}, \citenamefont
  {{Damascelli}}, \citenamefont {{van der Marel}},\ and\ \citenamefont
  {{Parmigiani}}}]{giannetti2011}%
  \BibitemOpen
  \bibfield  {author} {\bibinfo {author} {\bibfnamefont {C.}~\bibnamefont
  {{Giannetti}}}, \bibinfo {author} {\bibfnamefont {F.}~\bibnamefont
  {{Cilento}}}, \bibinfo {author} {\bibfnamefont {S.~D.}\ \bibnamefont
  {{Conte}}}, \bibinfo {author} {\bibfnamefont {G.}~\bibnamefont
  {{Coslovich}}}, \bibinfo {author} {\bibfnamefont {G.}~\bibnamefont
  {{Ferrini}}}, \bibinfo {author} {\bibfnamefont {H.}~\bibnamefont
  {{Molegraaf}}}, \bibinfo {author} {\bibfnamefont {M.}~\bibnamefont
  {{Raichle}}}, \bibinfo {author} {\bibfnamefont {R.}~\bibnamefont {{Liang}}},
  \bibinfo {author} {\bibfnamefont {H.}~\bibnamefont {{Eisaki}}}, \bibinfo
  {author} {\bibfnamefont {M.}~\bibnamefont {{Greven}}}, \bibinfo {author}
  {\bibfnamefont {A.}~\bibnamefont {{Damascelli}}}, \bibinfo {author}
  {\bibfnamefont {D.}~\bibnamefont {{van der Marel}}}, \ and\ \bibinfo {author}
  {\bibfnamefont {F.}~\bibnamefont {{Parmigiani}}},\ }\href {\doibase
  10.1038/ncomms1354} {\bibfield  {journal} {\bibinfo  {journal} {Nature
  Communications}\ }\textbf {\bibinfo {volume} {2}},\ \bibinfo {eid} {353}
  (\bibinfo {year} {2011})}\BibitemShut {NoStop}%
\bibitem [{\citenamefont {Zhang}\ and\ \citenamefont
  {Rice}(1988)}]{ZhangRice1988}%
  \BibitemOpen
  \bibfield  {author} {\bibinfo {author} {\bibfnamefont {F.~C.}\ \bibnamefont
  {Zhang}}\ and\ \bibinfo {author} {\bibfnamefont {T.~M.}\ \bibnamefont
  {Rice}},\ }\href {\doibase 10.1103/PhysRevB.37.3759} {\bibfield  {journal}
  {\bibinfo  {journal} {Phys. Rev. B}\ }\textbf {\bibinfo {volume} {37}},\
  \bibinfo {pages} {3759} (\bibinfo {year} {1988})}\BibitemShut {NoStop}%
\bibitem [{\citenamefont {Kohsaka}\ \emph {et~al.}(2007)\citenamefont
  {Kohsaka}, \citenamefont {Taylor}, \citenamefont {Fujita}, \citenamefont
  {Schmidt}, \citenamefont {Lupien}, \citenamefont {Hanaguri}, \citenamefont
  {Azuma}, \citenamefont {Takano}, \citenamefont {Eisaki}, \citenamefont
  {Takagi}, \citenamefont {Uchida},\ and\ \citenamefont {Davis}}]{Davis:2007}%
  \BibitemOpen
  \bibfield  {author} {\bibinfo {author} {\bibfnamefont {Y.}~\bibnamefont
  {Kohsaka}}, \bibinfo {author} {\bibfnamefont {C.}~\bibnamefont {Taylor}},
  \bibinfo {author} {\bibfnamefont {K.}~\bibnamefont {Fujita}}, \bibinfo
  {author} {\bibfnamefont {A.}~\bibnamefont {Schmidt}}, \bibinfo {author}
  {\bibfnamefont {C.}~\bibnamefont {Lupien}}, \bibinfo {author} {\bibfnamefont
  {T.}~\bibnamefont {Hanaguri}}, \bibinfo {author} {\bibfnamefont
  {M.}~\bibnamefont {Azuma}}, \bibinfo {author} {\bibfnamefont
  {M.}~\bibnamefont {Takano}}, \bibinfo {author} {\bibfnamefont
  {H.}~\bibnamefont {Eisaki}}, \bibinfo {author} {\bibfnamefont
  {H.}~\bibnamefont {Takagi}}, \bibinfo {author} {\bibfnamefont
  {S.}~\bibnamefont {Uchida}}, \ and\ \bibinfo {author} {\bibfnamefont {J.~C.}\
  \bibnamefont {Davis}},\ }\href {\doibase 10.1126/science.1138584} {\bibfield
  {journal} {\bibinfo  {journal} {Science}\ }\textbf {\bibinfo {volume}
  {315}},\ \bibinfo {pages} {1380} (\bibinfo {year} {2007})}\BibitemShut
  {NoStop}%
\bibitem [{\citenamefont {Timusk}\ and\ \citenamefont {Statt}(1999)}]{timusk}%
  \BibitemOpen
  \bibfield  {author} {\bibinfo {author} {\bibfnamefont {T.}~\bibnamefont
  {Timusk}}\ and\ \bibinfo {author} {\bibfnamefont {B.}~\bibnamefont {Statt}},\
  }\href {http://stacks.iop.org/0034-4885/62/i=1/a=002} {\bibfield  {journal}
  {\bibinfo  {journal} {Reports on Progress in Physics}\ }\textbf {\bibinfo
  {volume} {62}},\ \bibinfo {pages} {61} (\bibinfo {year} {1999})}\BibitemShut
  {NoStop}%
\bibitem [{\citenamefont {Chubukov}\ \emph {et~al.}(2008)\citenamefont
  {Chubukov}, \citenamefont {Pines},\ and\ \citenamefont
  {Schmalian}}]{Chubukov2008}%
  \BibitemOpen
  \bibfield  {author} {\bibinfo {author} {\bibfnamefont {A.~V.}\ \bibnamefont
  {Chubukov}}, \bibinfo {author} {\bibfnamefont {D.}~\bibnamefont {Pines}}, \
  and\ \bibinfo {author} {\bibfnamefont {J.}~\bibnamefont {Schmalian}},\
  }\enquote {\bibinfo {title} {Superconductivity: Conventional and
  unconventional superconductors},}\ \ (\bibinfo  {publisher} {Springer Berlin
  Heidelberg},\ \bibinfo {address} {Berlin, Heidelberg},\ \bibinfo {year}
  {2008})\ Chap.\ \bibinfo {chapter} {A Spin Fluctuation Model for d-Wave
  Superconductivity}, pp.\ \bibinfo {pages} {1349--1413}\BibitemShut {NoStop}%
\bibitem [{\citenamefont {Efetov}\ \emph {et~al.}(2013)\citenamefont {Efetov},
  \citenamefont {Meier},\ and\ \citenamefont {Pepin}}]{efetov_pseudogap_2013}%
  \BibitemOpen
  \bibfield  {author} {\bibinfo {author} {\bibfnamefont {K.~B.}\ \bibnamefont
  {Efetov}}, \bibinfo {author} {\bibfnamefont {H.}~\bibnamefont {Meier}}, \
  and\ \bibinfo {author} {\bibfnamefont {C.}~\bibnamefont {Pepin}},\ }\href
  {http://dx.doi.org/10.1038/nphys2641} {\bibfield  {journal} {\bibinfo
  {journal} {Nat Phys}\ }\textbf {\bibinfo {volume} {9}},\ \bibinfo {pages}
  {442} (\bibinfo {year} {2013})}\BibitemShut {NoStop}%
\bibitem [{\citenamefont {Sachdev}\ and\ \citenamefont
  {La~Placa}(2013)}]{SachedevBond2013}%
  \BibitemOpen
  \bibfield  {author} {\bibinfo {author} {\bibfnamefont {S.}~\bibnamefont
  {Sachdev}}\ and\ \bibinfo {author} {\bibfnamefont {R.}~\bibnamefont
  {La~Placa}},\ }\href {\doibase 10.1103/PhysRevLett.111.027202} {\bibfield
  {journal} {\bibinfo  {journal} {Phys. Rev. Lett.}\ }\textbf {\bibinfo
  {volume} {111}},\ \bibinfo {pages} {027202} (\bibinfo {year}
  {2013})}\BibitemShut {NoStop}%
\bibitem [{\citenamefont {Wang}\ and\ \citenamefont
  {Chubukov}(2014)}]{ChubukovCDW2014}%
  \BibitemOpen
  \bibfield  {author} {\bibinfo {author} {\bibfnamefont {Y.}~\bibnamefont
  {Wang}}\ and\ \bibinfo {author} {\bibfnamefont {A.}~\bibnamefont
  {Chubukov}},\ }\href {\doibase 10.1103/PhysRevB.90.035149} {\bibfield
  {journal} {\bibinfo  {journal} {Phys. Rev. B}\ }\textbf {\bibinfo {volume}
  {90}},\ \bibinfo {pages} {035149} (\bibinfo {year} {2014})}\BibitemShut
  {NoStop}%
\bibitem [{\citenamefont {Varma}\ \emph {et~al.}(1989)\citenamefont {Varma},
  \citenamefont {Littlewood}, \citenamefont {Schmitt-Rink}, \citenamefont
  {Abrahams},\ and\ \citenamefont {Ruckenstein}}]{Varma1989}%
  \BibitemOpen
  \bibfield  {author} {\bibinfo {author} {\bibfnamefont {C.~M.}\ \bibnamefont
  {Varma}}, \bibinfo {author} {\bibfnamefont {P.~B.}\ \bibnamefont
  {Littlewood}}, \bibinfo {author} {\bibfnamefont {S.}~\bibnamefont
  {Schmitt-Rink}}, \bibinfo {author} {\bibfnamefont {E.}~\bibnamefont
  {Abrahams}}, \ and\ \bibinfo {author} {\bibfnamefont {A.~E.}\ \bibnamefont
  {Ruckenstein}},\ }\href {\doibase 10.1103/PhysRevLett.63.1996} {\bibfield
  {journal} {\bibinfo  {journal} {Phys. Rev. Lett.}\ }\textbf {\bibinfo
  {volume} {63}},\ \bibinfo {pages} {1996} (\bibinfo {year}
  {1989})}\BibitemShut {NoStop}%
\bibitem [{\citenamefont {Varma}(2006)}]{Varma2006}%
  \BibitemOpen
  \bibfield  {author} {\bibinfo {author} {\bibfnamefont {C.~M.}\ \bibnamefont
  {Varma}},\ }\href {\doibase 10.1103/PhysRevB.73.155113} {\bibfield  {journal}
  {\bibinfo  {journal} {Phys. Rev. B}\ }\textbf {\bibinfo {volume} {73}},\
  \bibinfo {pages} {155113} (\bibinfo {year} {2006})}\BibitemShut {NoStop}%
\bibitem [{\citenamefont {Weber}\ \emph {et~al.}(2014)\citenamefont {Weber},
  \citenamefont {Giamarchi},\ and\ \citenamefont {Varma}}]{Cedric2014}%
  \BibitemOpen
  \bibfield  {author} {\bibinfo {author} {\bibfnamefont {C.}~\bibnamefont
  {Weber}}, \bibinfo {author} {\bibfnamefont {T.}~\bibnamefont {Giamarchi}}, \
  and\ \bibinfo {author} {\bibfnamefont {C.~M.}\ \bibnamefont {Varma}},\ }\href
  {\doibase 10.1103/PhysRevLett.112.117001} {\bibfield  {journal} {\bibinfo
  {journal} {Phys. Rev. Lett.}\ }\textbf {\bibinfo {volume} {112}},\ \bibinfo
  {pages} {117001} (\bibinfo {year} {2014})}\BibitemShut {NoStop}%
\bibitem [{\citenamefont {Bulut}\ \emph {et~al.}(2015)\citenamefont {Bulut},
  \citenamefont {Kampf},\ and\ \citenamefont {Atkinson}}]{Bulut2014}%
  \BibitemOpen
  \bibfield  {author} {\bibinfo {author} {\bibfnamefont {S.}~\bibnamefont
  {Bulut}}, \bibinfo {author} {\bibfnamefont {A.~P.}\ \bibnamefont {Kampf}}, \
  and\ \bibinfo {author} {\bibfnamefont {W.~A.}\ \bibnamefont {Atkinson}},\
  }\href {\doibase 10.1103/PhysRevB.92.195140} {\bibfield  {journal} {\bibinfo
  {journal} {Phys. Rev. B}\ }\textbf {\bibinfo {volume} {92}},\ \bibinfo
  {pages} {195140} (\bibinfo {year} {2015})}\BibitemShut {NoStop}%
\bibitem [{\citenamefont {Kung}\ \emph {et~al.}(2014)\citenamefont {Kung},
  \citenamefont {Chen}, \citenamefont {Moritz}, \citenamefont {Johnston},
  \citenamefont {Thomale},\ and\ \citenamefont {Devereaux}}]{Kung2014}%
  \BibitemOpen
  \bibfield  {author} {\bibinfo {author} {\bibfnamefont {Y.~F.}\ \bibnamefont
  {Kung}}, \bibinfo {author} {\bibfnamefont {C.-C.}\ \bibnamefont {Chen}},
  \bibinfo {author} {\bibfnamefont {B.}~\bibnamefont {Moritz}}, \bibinfo
  {author} {\bibfnamefont {S.}~\bibnamefont {Johnston}}, \bibinfo {author}
  {\bibfnamefont {R.}~\bibnamefont {Thomale}}, \ and\ \bibinfo {author}
  {\bibfnamefont {T.~P.}\ \bibnamefont {Devereaux}},\ }\href {\doibase
  10.1103/PhysRevB.90.224507} {\bibfield  {journal} {\bibinfo  {journal} {Phys.
  Rev. B}\ }\textbf {\bibinfo {volume} {90}},\ \bibinfo {pages} {224507}
  (\bibinfo {year} {2014})}\BibitemShut {NoStop}%
\bibitem [{\citenamefont {de~Carvalho}\ \emph {et~al.}(2016)\citenamefont
  {de~Carvalho}, \citenamefont {P\'epin},\ and\ \citenamefont
  {Freire}}]{Pepin2016}%
  \BibitemOpen
  \bibfield  {author} {\bibinfo {author} {\bibfnamefont {V.~S.}\ \bibnamefont
  {de~Carvalho}}, \bibinfo {author} {\bibfnamefont {C.}~\bibnamefont
  {P\'epin}}, \ and\ \bibinfo {author} {\bibfnamefont {H.}~\bibnamefont
  {Freire}},\ }\href {\doibase 10.1103/PhysRevB.93.115144} {\bibfield
  {journal} {\bibinfo  {journal} {Phys. Rev. B}\ }\textbf {\bibinfo {volume}
  {93}},\ \bibinfo {pages} {115144} (\bibinfo {year} {2016})}\BibitemShut
  {NoStop}%
\bibitem [{\citenamefont {Sordi}\ \emph {et~al.}(2010)\citenamefont {Sordi},
  \citenamefont {Haule},\ and\ \citenamefont {Tremblay}}]{sht}%
  \BibitemOpen
  \bibfield  {author} {\bibinfo {author} {\bibfnamefont {G.}~\bibnamefont
  {Sordi}}, \bibinfo {author} {\bibfnamefont {K.}~\bibnamefont {Haule}}, \ and\
  \bibinfo {author} {\bibfnamefont {A.-M.~S.}\ \bibnamefont {Tremblay}},\
  }\href {\doibase 10.1103/PhysRevLett.104.226402} {\bibfield  {journal}
  {\bibinfo  {journal} {Phys. Rev. Lett.}\ }\textbf {\bibinfo {volume} {104}},\
  \bibinfo {pages} {226402} (\bibinfo {year} {2010})}\BibitemShut {NoStop}%
\bibitem [{\citenamefont {Sordi}\ \emph {et~al.}(2011)\citenamefont {Sordi},
  \citenamefont {Haule},\ and\ \citenamefont {Tremblay}}]{sht2}%
  \BibitemOpen
  \bibfield  {author} {\bibinfo {author} {\bibfnamefont {G.}~\bibnamefont
  {Sordi}}, \bibinfo {author} {\bibfnamefont {K.}~\bibnamefont {Haule}}, \ and\
  \bibinfo {author} {\bibfnamefont {A.-M.~S.}\ \bibnamefont {Tremblay}},\
  }\href {\doibase 10.1103/PhysRevB.84.075161} {\bibfield  {journal} {\bibinfo
  {journal} {Phys. Rev. B}\ }\textbf {\bibinfo {volume} {84}},\ \bibinfo
  {pages} {075161} (\bibinfo {year} {2011})}\BibitemShut {NoStop}%
\bibitem [{\citenamefont {Sordi}\ \emph
  {et~al.}(2012{\natexlab{b}})\citenamefont {Sordi}, \citenamefont {S\'emon},
  \citenamefont {Haule},\ and\ \citenamefont {Tremblay}}]{sshtSC}%
  \BibitemOpen
  \bibfield  {author} {\bibinfo {author} {\bibfnamefont {G.}~\bibnamefont
  {Sordi}}, \bibinfo {author} {\bibfnamefont {P.}~\bibnamefont {S\'emon}},
  \bibinfo {author} {\bibfnamefont {K.}~\bibnamefont {Haule}}, \ and\ \bibinfo
  {author} {\bibfnamefont {A.-M.~S.}\ \bibnamefont {Tremblay}},\ }\href
  {\doibase 10.1103/PhysRevLett.108.216401} {\bibfield  {journal} {\bibinfo
  {journal} {Phys. Rev. Lett.}\ }\textbf {\bibinfo {volume} {108}},\ \bibinfo
  {pages} {216401} (\bibinfo {year} {2012}{\natexlab{b}})}\BibitemShut
  {NoStop}%
\bibitem [{\citenamefont {Sordi}\ \emph {et~al.}(2013)\citenamefont {Sordi},
  \citenamefont {S\'emon}, \citenamefont {Haule},\ and\ \citenamefont
  {Tremblay}}]{sshtRHO}%
  \BibitemOpen
  \bibfield  {author} {\bibinfo {author} {\bibfnamefont {G.}~\bibnamefont
  {Sordi}}, \bibinfo {author} {\bibfnamefont {P.}~\bibnamefont {S\'emon}},
  \bibinfo {author} {\bibfnamefont {K.}~\bibnamefont {Haule}}, \ and\ \bibinfo
  {author} {\bibfnamefont {A.-M.~S.}\ \bibnamefont {Tremblay}},\ }\href
  {\doibase 10.1103/PhysRevB.87.041101} {\bibfield  {journal} {\bibinfo
  {journal} {Phys. Rev. B}\ }\textbf {\bibinfo {volume} {87}},\ \bibinfo
  {pages} {041101} (\bibinfo {year} {2013})}\BibitemShut {NoStop}%
\bibitem [{\citenamefont {Fratino}\ \emph {et~al.}(2016)\citenamefont
  {Fratino}, \citenamefont {S\'emon}, \citenamefont {Sordi},\ and\
  \citenamefont {Tremblay}}]{LorenzoSC}%
  \BibitemOpen
  \bibfield  {author} {\bibinfo {author} {\bibfnamefont {L.}~\bibnamefont
  {Fratino}}, \bibinfo {author} {\bibfnamefont {P.}~\bibnamefont {S\'emon}},
  \bibinfo {author} {\bibfnamefont {G.}~\bibnamefont {Sordi}}, \ and\ \bibinfo
  {author} {\bibfnamefont {A.-M.~S.}\ \bibnamefont {Tremblay}},\ }\href
  {\doibase 10.1038/srep22715} {\bibfield  {journal} {\bibinfo  {journal} {Sci.
  Rep.}\ }\textbf {\bibinfo {volume} {6}},\ \bibinfo {pages} {22715} (\bibinfo
  {year} {2016})}\BibitemShut {NoStop}%
\bibitem [{\citenamefont {Furukawa}\ \emph {et~al.}(2015)\citenamefont
  {Furukawa}, \citenamefont {Miyagawa}, \citenamefont {Taniguchi},
  \citenamefont {Kato},\ and\ \citenamefont {Kanoda}}]{furukawaWL}%
  \BibitemOpen
  \bibfield  {author} {\bibinfo {author} {\bibfnamefont {T.}~\bibnamefont
  {Furukawa}}, \bibinfo {author} {\bibfnamefont {K.}~\bibnamefont {Miyagawa}},
  \bibinfo {author} {\bibfnamefont {H.}~\bibnamefont {Taniguchi}}, \bibinfo
  {author} {\bibfnamefont {R.}~\bibnamefont {Kato}}, \ and\ \bibinfo {author}
  {\bibfnamefont {K.}~\bibnamefont {Kanoda}},\ }\href {\doibase
  10.1038/nphys3235} {\bibfield  {journal} {\bibinfo  {journal} {Nature
  Physics}\ }\textbf {\bibinfo {volume} {3}},\ \bibinfo {eid} {221} (\bibinfo
  {year} {2015})}\BibitemShut {NoStop}%
\bibitem [{\citenamefont {Terletska}\ \emph {et~al.}(2011)\citenamefont
  {Terletska}, \citenamefont {Vu\ifmmode \check{c}\else \v{c}\fi{}i\ifmmode
  \check{c}\else \v{c}\fi{}evi\ifmmode~\acute{c}\else \'{c}\fi{}},
  \citenamefont {Tanaskovi\ifmmode~\acute{c}\else \'{c}\fi{}},\ and\
  \citenamefont {Dobrosavljevi\ifmmode~\acute{c}\else \'{c}\fi{}}}]{vlad1}%
  \BibitemOpen
  \bibfield  {author} {\bibinfo {author} {\bibfnamefont {H.}~\bibnamefont
  {Terletska}}, \bibinfo {author} {\bibfnamefont {J.}~\bibnamefont {Vu\ifmmode
  \check{c}\else \v{c}\fi{}i\ifmmode \check{c}\else
  \v{c}\fi{}evi\ifmmode~\acute{c}\else \'{c}\fi{}}}, \bibinfo {author}
  {\bibfnamefont {D.}~\bibnamefont {Tanaskovi\ifmmode~\acute{c}\else
  \'{c}\fi{}}}, \ and\ \bibinfo {author} {\bibfnamefont {V.}~\bibnamefont
  {Dobrosavljevi\ifmmode~\acute{c}\else \'{c}\fi{}}},\ }\href {\doibase
  10.1103/PhysRevLett.107.026401} {\bibfield  {journal} {\bibinfo  {journal}
  {Phys. Rev. Lett.}\ }\textbf {\bibinfo {volume} {107}},\ \bibinfo {pages}
  {026401} (\bibinfo {year} {2011})}\BibitemShut {NoStop}%
\bibitem [{\citenamefont {Vu\ifmmode \check{c}\else \v{c}\fi{}i\ifmmode
  \check{c}\else \v{c}\fi{}evi\ifmmode~\acute{c}\else \'{c}\fi{}}\ \emph
  {et~al.}(2013)\citenamefont {Vu\ifmmode \check{c}\else \v{c}\fi{}i\ifmmode
  \check{c}\else \v{c}\fi{}evi\ifmmode~\acute{c}\else \'{c}\fi{}},
  \citenamefont {Terletska}, \citenamefont {Tanaskovi\ifmmode~\acute{c}\else
  \'{c}\fi{}},\ and\ \citenamefont {Dobrosavljevi\ifmmode~\acute{c}\else
  \'{c}\fi{}}}]{vlad2}%
  \BibitemOpen
  \bibfield  {author} {\bibinfo {author} {\bibfnamefont {J.}~\bibnamefont
  {Vu\ifmmode \check{c}\else \v{c}\fi{}i\ifmmode \check{c}\else
  \v{c}\fi{}evi\ifmmode~\acute{c}\else \'{c}\fi{}}}, \bibinfo {author}
  {\bibfnamefont {H.}~\bibnamefont {Terletska}}, \bibinfo {author}
  {\bibfnamefont {D.}~\bibnamefont {Tanaskovi\ifmmode~\acute{c}\else
  \'{c}\fi{}}}, \ and\ \bibinfo {author} {\bibfnamefont {V.}~\bibnamefont
  {Dobrosavljevi\ifmmode~\acute{c}\else \'{c}\fi{}}},\ }\href {\doibase
  10.1103/PhysRevB.88.075143} {\bibfield  {journal} {\bibinfo  {journal} {Phys.
  Rev. B}\ }\textbf {\bibinfo {volume} {88}},\ \bibinfo {pages} {075143}
  (\bibinfo {year} {2013})}\BibitemShut {NoStop}%
\bibitem [{\citenamefont {Vu\ifmmode \check{c}\else \v{c}\fi{}i\ifmmode
  \check{c}\else \v{c}\fi{}evi\ifmmode~\acute{c}\else \'{c}\fi{}}\ \emph
  {et~al.}(2015)\citenamefont {Vu\ifmmode \check{c}\else \v{c}\fi{}i\ifmmode
  \check{c}\else \v{c}\fi{}evi\ifmmode~\acute{c}\else \'{c}\fi{}},
  \citenamefont {Tanaskovi\ifmmode~\acute{c}\else \'{c}\fi{}}, \citenamefont
  {Rozenberg},\ and\ \citenamefont {Dobrosavljevi\ifmmode~\acute{c}\else
  \'{c}\fi{}}}]{vlad3}%
  \BibitemOpen
  \bibfield  {author} {\bibinfo {author} {\bibfnamefont {J.}~\bibnamefont
  {Vu\ifmmode \check{c}\else \v{c}\fi{}i\ifmmode \check{c}\else
  \v{c}\fi{}evi\ifmmode~\acute{c}\else \'{c}\fi{}}}, \bibinfo {author}
  {\bibfnamefont {D.}~\bibnamefont {Tanaskovi\ifmmode~\acute{c}\else
  \'{c}\fi{}}}, \bibinfo {author} {\bibfnamefont {M.~J.}\ \bibnamefont
  {Rozenberg}}, \ and\ \bibinfo {author} {\bibfnamefont {V.}~\bibnamefont
  {Dobrosavljevi\ifmmode~\acute{c}\else \'{c}\fi{}}},\ }\href {\doibase
  10.1103/PhysRevLett.114.246402} {\bibfield  {journal} {\bibinfo  {journal}
  {Phys. Rev. Lett.}\ }\textbf {\bibinfo {volume} {114}},\ \bibinfo {pages}
  {246402} (\bibinfo {year} {2015})}\BibitemShut {NoStop}%
\bibitem [{\citenamefont {H\'ebert}\ \emph {et~al.}(2015)\citenamefont
  {H\'ebert}, \citenamefont {S\'emon},\ and\ \citenamefont
  {Tremblay}}]{Hebert:2015}%
  \BibitemOpen
  \bibfield  {author} {\bibinfo {author} {\bibfnamefont {C.-D.}\ \bibnamefont
  {H\'ebert}}, \bibinfo {author} {\bibfnamefont {P.}~\bibnamefont {S\'emon}}, \
  and\ \bibinfo {author} {\bibfnamefont {A.-M.~S.}\ \bibnamefont {Tremblay}},\
  }\href {\doibase 10.1103/PhysRevB.92.195112} {\bibfield  {journal} {\bibinfo
  {journal} {Phys. Rev. B}\ }\textbf {\bibinfo {volume} {92}},\ \bibinfo
  {pages} {195112} (\bibinfo {year} {2015})}\BibitemShut {NoStop}%
\bibitem [{\citenamefont {McMillan}\ and\ \citenamefont
  {Stanley}(2010)}]{supercritical}%
  \BibitemOpen
  \bibfield  {author} {\bibinfo {author} {\bibfnamefont {P.~F.}\ \bibnamefont
  {McMillan}}\ and\ \bibinfo {author} {\bibfnamefont {H.~E.}\ \bibnamefont
  {Stanley}},\ }\href {\doibase doi:10.1038/nphys1711} {\bibfield  {journal}
  {\bibinfo  {journal} {Nat Phys}\ }\textbf {\bibinfo {volume} {6}},\ \bibinfo
  {pages} {479} (\bibinfo {year} {2010})}\BibitemShut {NoStop}%
\bibitem [{\citenamefont {Simeoni}\ \emph {et~al.}(2010)\citenamefont
  {Simeoni}, \citenamefont {Bryk}, \citenamefont {Gorelli}, \citenamefont
  {Krisch}, \citenamefont {Ruocco}, \citenamefont {Santoro},\ and\
  \citenamefont {Scopigno}}]{Simeoni2010}%
  \BibitemOpen
  \bibfield  {author} {\bibinfo {author} {\bibfnamefont {G.~G.}\ \bibnamefont
  {Simeoni}}, \bibinfo {author} {\bibfnamefont {T.}~\bibnamefont {Bryk}},
  \bibinfo {author} {\bibfnamefont {F.~A.}\ \bibnamefont {Gorelli}}, \bibinfo
  {author} {\bibfnamefont {M.}~\bibnamefont {Krisch}}, \bibinfo {author}
  {\bibfnamefont {G.}~\bibnamefont {Ruocco}}, \bibinfo {author} {\bibfnamefont
  {M.}~\bibnamefont {Santoro}}, \ and\ \bibinfo {author} {\bibfnamefont
  {T.}~\bibnamefont {Scopigno}},\ }\href {\doibase doi:10.1038/nphys1683}
  {\bibfield  {journal} {\bibinfo  {journal} {Nature Physics}\ }\textbf
  {\bibinfo {volume} {6}},\ \bibinfo {pages} {503} (\bibinfo {year}
  {2010})}\BibitemShut {NoStop}%
\bibitem [{\citenamefont {Lacey}(2015)}]{Lacey2015}%
  \BibitemOpen
  \bibfield  {author} {\bibinfo {author} {\bibfnamefont {R.~A.}\ \bibnamefont
  {Lacey}},\ }\href {\doibase 10.1103/PhysRevLett.114.142301} {\bibfield
  {journal} {\bibinfo  {journal} {Phys. Rev. Lett.}\ }\textbf {\bibinfo
  {volume} {114}},\ \bibinfo {pages} {142301} (\bibinfo {year}
  {2015})}\BibitemShut {NoStop}%
\bibitem [{\citenamefont {Stephanov}\ \emph {et~al.}(1998)\citenamefont
  {Stephanov}, \citenamefont {Rajagopal},\ and\ \citenamefont
  {Shuryak}}]{Stephanov1998}%
  \BibitemOpen
  \bibfield  {author} {\bibinfo {author} {\bibfnamefont {M.}~\bibnamefont
  {Stephanov}}, \bibinfo {author} {\bibfnamefont {K.}~\bibnamefont
  {Rajagopal}}, \ and\ \bibinfo {author} {\bibfnamefont {E.}~\bibnamefont
  {Shuryak}},\ }\href {\doibase 10.1103/PhysRevLett.81.4816} {\bibfield
  {journal} {\bibinfo  {journal} {Phys. Rev. Lett.}\ }\textbf {\bibinfo
  {volume} {81}},\ \bibinfo {pages} {4816} (\bibinfo {year}
  {1998})}\BibitemShut {NoStop}%
\bibitem [{\citenamefont {Stephanov}(2004)}]{StephanovRev}%
  \BibitemOpen
  \bibfield  {author} {\bibinfo {author} {\bibfnamefont {M.}~\bibnamefont
  {Stephanov}},\ }\href {\doibase 10.1143/PTPS.153.139} {\bibfield  {journal}
  {\bibinfo  {journal} {Prog. Theor. Phys. Supplement}\ }\textbf {\bibinfo
  {volume} {153}},\ \bibinfo {pages} {139} (\bibinfo {year}
  {2004})}\BibitemShut {NoStop}%
\end{thebibliography}

\clearpage
\begin{thebibliography}{11}

\makeatletter
\providecommand \@ifxundefined [1]{%
 \@ifx{#1\undefined}
}%
\providecommand \@ifnum [1]{%
 \ifnum #1\expandafter \@firstoftwo
 \else \expandafter \@secondoftwo
 \fi
}%
\providecommand \@ifx [1]{%
 \ifx #1\expandafter \@firstoftwo
 \else \expandafter \@secondoftwo
 \fi
}%
\providecommand \natexlab [1]{#1}%
\providecommand \enquote  [1]{``#1''}%
\providecommand \bibnamefont  [1]{#1}%
\providecommand \bibfnamefont [1]{#1}%
\providecommand \citenamefont [1]{#1}%
\providecommand \href@noop [0]{\@secondoftwo}%
\providecommand \href [0]{\begingroup \@sanitize@url \@href}%
\providecommand \@href[1]{\@@startlink{#1}\@@href}%
\providecommand \@@href[1]{\endgroup#1\@@endlink}%
\providecommand \@sanitize@url [0]{\catcode `\\12\catcode `\$12\catcode
  `\&12\catcode `\#12\catcode `\^12\catcode `\_12\catcode `\%12\relax}%
\providecommand \@@startlink[1]{}%
\providecommand \@@endlink[0]{}%
\providecommand \url  [0]{\begingroup\@sanitize@url \@url }%
\providecommand \@url [1]{\endgroup\@href {#1}{\urlprefix }}%
\providecommand \urlprefix  [0]{URL }%
\providecommand \Eprint [0]{\href }%
\providecommand \doibase [0]{http://dx.doi.org/}%
\providecommand \selectlanguage [0]{\@gobble}%
\providecommand \bibinfo  [0]{\@secondoftwo}%
\providecommand \bibfield  [0]{\@secondoftwo}%
\providecommand \translation [1]{[#1]}%
\providecommand \BibitemOpen [0]{}%
\providecommand \bibitemStop [0]{}%
\providecommand \bibitemNoStop [0]{.\EOS\space}%
\providecommand \EOS [0]{\spacefactor3000\relax}%
\providecommand \BibitemShut  [1]{\csname bibitem#1\endcsname}%
\let\auto@bib@innerbib\@empty
\bibitem [{\citenamefont {Maier}\ \emph {et~al.}(2005)\citenamefont {Maier},
  \citenamefont {Jarrell}, \citenamefont {Pruschke},\ and\ \citenamefont
  {Hettler}}]{maier_SM}%
  \BibitemOpen
  \bibfield  {author} {\bibinfo {author} {\bibfnamefont {T.}~\bibnamefont
  {Maier}}, \bibinfo {author} {\bibfnamefont {M.}~\bibnamefont {Jarrell}},
  \bibinfo {author} {\bibfnamefont {T.}~\bibnamefont {Pruschke}}, \ and\
  \bibinfo {author} {\bibfnamefont {M.~H.}\ \bibnamefont {Hettler}},\ }\href
  {\doibase 10.1103/RevModPhys.77.1027} {\bibfield  {journal} {\bibinfo
  {journal} {Rev. Mod. Phys.}\ }\textbf {\bibinfo {volume} {77}},\ \bibinfo
  {pages} {1027} (\bibinfo {year} {2005})}\BibitemShut {NoStop}%
\bibitem [{\citenamefont {Kotliar}\ \emph {et~al.}(2006)\citenamefont
  {Kotliar}, \citenamefont {Savrasov}, \citenamefont {Haule}, \citenamefont
  {Oudovenko}, \citenamefont {Parcollet},\ and\ \citenamefont
  {Marianetti}}]{kotliarRMP_SM}%
  \BibitemOpen
  \bibfield  {author} {\bibinfo {author} {\bibfnamefont {G.}~\bibnamefont
  {Kotliar}}, \bibinfo {author} {\bibfnamefont {S.~Y.}\ \bibnamefont
  {Savrasov}}, \bibinfo {author} {\bibfnamefont {K.}~\bibnamefont {Haule}},
  \bibinfo {author} {\bibfnamefont {V.~S.}\ \bibnamefont {Oudovenko}}, \bibinfo
  {author} {\bibfnamefont {O.}~\bibnamefont {Parcollet}}, \ and\ \bibinfo
  {author} {\bibfnamefont {C.~A.}\ \bibnamefont {Marianetti}},\ }\href
  {\doibase 10.1103/RevModPhys.78.865} {\bibfield  {journal} {\bibinfo
  {journal} {Rev. Mod. Phys.}\ }\textbf {\bibinfo {volume} {78}},\ \bibinfo
  {eid} {865} (\bibinfo {year} {2006})}\BibitemShut {NoStop}%
\bibitem [{\citenamefont {Tremblay}\ \emph {et~al.}(2006)\citenamefont
  {Tremblay}, \citenamefont {Kyung},\ and\ \citenamefont
  {S\'{e}n\'{e}chal}}]{tremblayR_SM}%
  \BibitemOpen
  \bibfield  {author} {\bibinfo {author} {\bibfnamefont {A.-M.~S.}\
  \bibnamefont {Tremblay}}, \bibinfo {author} {\bibfnamefont {B.}~\bibnamefont
  {Kyung}}, \ and\ \bibinfo {author} {\bibfnamefont {D.}~\bibnamefont
  {S\'{e}n\'{e}chal}},\ }\href {\doibase 10.1063/1.2199446} {\bibfield
  {journal} {\bibinfo  {journal} {Low Temp. Phys.}\ }\textbf {\bibinfo {volume}
  {32}},\ \bibinfo {pages} {424} (\bibinfo {year} {2006})}\BibitemShut
  {NoStop}%
\bibitem [{\citenamefont {Gull}\ \emph {et~al.}(2011)\citenamefont {Gull},
  \citenamefont {Millis}, \citenamefont {Lichtenstein}, \citenamefont
  {Rubtsov}, \citenamefont {Troyer},\ and\ \citenamefont {Werner}}]{millisRMP_SM}%
  \BibitemOpen
  \bibfield  {author} {\bibinfo {author} {\bibfnamefont {E.}~\bibnamefont
  {Gull}}, \bibinfo {author} {\bibfnamefont {A.~J.}\ \bibnamefont {Millis}},
  \bibinfo {author} {\bibfnamefont {A.~I.}\ \bibnamefont {Lichtenstein}},
  \bibinfo {author} {\bibfnamefont {A.~N.}\ \bibnamefont {Rubtsov}}, \bibinfo
  {author} {\bibfnamefont {M.}~\bibnamefont {Troyer}}, \ and\ \bibinfo {author}
  {\bibfnamefont {P.}~\bibnamefont {Werner}},\ }\href {\doibase
  10.1103/RevModPhys.83.349} {\bibfield  {journal} {\bibinfo  {journal} {Rev.
  Mod. Phys.}\ }\textbf {\bibinfo {volume} {83}},\ \bibinfo {pages} {349}
  (\bibinfo {year} {2011})}\BibitemShut {NoStop}%
\bibitem [{\citenamefont {Andersen}\ \emph {et~al.}(1995)\citenamefont
  {Andersen}, \citenamefont {Liechtenstein}, \citenamefont {Jepsen},\ and\
  \citenamefont {Paulsen}}]{AndersenLDA_SM}%
  \BibitemOpen
  \bibfield  {author} {\bibinfo {author} {\bibfnamefont {O.}~\bibnamefont
  {Andersen}}, \bibinfo {author} {\bibfnamefont {A.}~\bibnamefont
  {Liechtenstein}}, \bibinfo {author} {\bibfnamefont {O.}~\bibnamefont
  {Jepsen}}, \ and\ \bibinfo {author} {\bibfnamefont {F.}~\bibnamefont
  {Paulsen}},\ }\href {\doibase http://dx.doi.org/10.1016/0022-3697(95)00269-3}
  {\bibfield  {journal} {\bibinfo  {journal} {J. Phys, Chem. Solids}\ }\textbf
  {\bibinfo {volume} {56}},\ \bibinfo {pages} {1573} (\bibinfo {year}
  {1995})}\BibitemShut {NoStop}%
\bibitem [{\citenamefont {Zaanen}\ \emph {et~al.}(1985)\citenamefont {Zaanen},
  \citenamefont {Sawatzky},\ and\ \citenamefont {Allen}}]{zsa_SM}%
  \BibitemOpen
  \bibfield  {author} {\bibinfo {author} {\bibfnamefont {J.}~\bibnamefont
  {Zaanen}}, \bibinfo {author} {\bibfnamefont {G.~A.}\ \bibnamefont
  {Sawatzky}}, \ and\ \bibinfo {author} {\bibfnamefont {J.~W.}\ \bibnamefont
  {Allen}},\ }\href {\doibase 10.1103/PhysRevLett.55.418} {\bibfield  {journal}
  {\bibinfo  {journal} {Phys. Rev. Lett.}\ }\textbf {\bibinfo {volume} {55}},\
  \bibinfo {pages} {418} (\bibinfo {year} {1985})}\BibitemShut {NoStop}%
\bibitem [{\citenamefont {{Bergeron}}\ and\ \citenamefont
  {{Tremblay}}(2015)}]{BergeronMaxEnt:2015_SM}%
  \BibitemOpen
  \bibfield  {author} {\bibinfo {author} {\bibfnamefont {D.}~\bibnamefont
  {{Bergeron}}}\ and\ \bibinfo {author} {\bibfnamefont {A.-M.~S.}\ \bibnamefont
  {{Tremblay}}},\ }\href {http://arxiv.org/abs/1507.01012} {\bibfield
  {journal} {\bibinfo  {journal} {ArXiv e-prints}\ } (\bibinfo {year}
  {2015})},\ \Eprint {http://arxiv.org/abs/1507.01012} {arXiv:1507.01012
  [cond-mat.str-el]} \BibitemShut {NoStop}%
\end{thebibliography}

\end{document}